\documentclass[%
 reprint,
superscriptaddress,
nofootinbib,
 amsmath,amssymb,
 aps,
 onecolumn,
]{revtex4-2}
\usepackage[utf8]{inputenc}
\usepackage{amsmath}
\usepackage{graphicx}
\usepackage{bm}
\usepackage{soul}
\usepackage{color}
\usepackage{subfig}

\newcommand{\eq}[1]{Eq.~(\ref{eq:#1})}
\newcommand{\qt}{{\vec q}_T}
\newcommand{\bt}{{\vec b}_T}
\newcommand{\refcite}[1]{Ref.~\cite{#1}}
\newcommand{\refscite}[1]{Refs.~\cite{#1}}
\newcommand{\Eq}[1]{Eq.~\eqref{eq:#1}}
\newcommand{\eqs}[2]{Eqs.~\eqref{eq:#1} and \eqref{eq:#2}}

\newcommand{\fig}[1]{Fig.~\ref{fig:#1}}

\newcommand{\cK}{\mathcal{K}}
\newcommand{\cO}{\mathcal{O}}
\newcommand{\cL}{\mathcal{L}}
\newcommand{\df}{\mathrm{d}}
\newcommand{\zb}{\bar{z}}
\newcommand{\img}{\mathrm{i}}
\def\beq{\begin{equation}}
\def\eeq{\end{equation}}
\newcommand{\born}{ \hat \sigma_0}
\newcommand{\as}{\alpha_s}
\newcommand{\GammaC}{\Gamma_{\rm cusp}}
\newcommand{\nn}{\nonumber}
\newcommand{\lqcd}{\Lambda_\mathrm{QCD}}
\newcommand{\zcut}{z_{/rm cut}}

\newcommand\snowmass{
\begin{center}
  \rule[-0.2in]{\hsize}{0.01in}\\
  \rule{\hsize}{0.01in}\\
  \vskip 0.1in
  Submitted to the Proceedings of the US Community Study\\ 
  on the Future of Particle Physics (Snowmass 2021)\\
  \rule{\hsize}{0.01in}\\
  \rule[+0.2in]{\hsize}{0.01in}\\[-2em]
\end{center}
}

\begin{document}

{
\begin{flushright}
SLAC-PUB-17665
\end{flushright}
}

\snowmass

\title{Energy-Energy Correlators for Precision QCD}

\author{Duff Neill}
\email{dneill@lanl.gov}
\affiliation{Los Alamos National Laboratory, Theoretical Division, Los Alamos, NM, 87545, USA}

\author{Gherardo Vita}
\email{gherardo@slac.stanford.edu}
\affiliation{SLAC National Accelerator Laboratory, Stanford University, Stanford, CA 94039, USA}

\author{Ivan Vitev}
\email{ivitev@lanl.gov}
\affiliation{Los Alamos National Laboratory, Theoretical Division, Los Alamos, NM, 87545, USA}

\author{Hua Xing Zhu}
\email{zhuhx@zju.edu.cn}
\affiliation{Zhejiang Institute of Modern Physics, Department of Physics, Zhejiang University, Hangzhou, 310027, China}

%\vspace*{1cm}
\begin{abstract}
  In this contribution to the Proceedings of the US Community Study
  on the Future of Particle Physics (Snowmass 2021) we review recent progress in the evaluation and application of the Energy-Energy Correlator (EEC) event shape observable in $e^+e^-$ annihilation, hadronic collisions, and deep inelastic scattering. The importance of EEC as a precision probe of the  perturbative and non perturbative aspects of QCD dynamics is emphasized. 
  %It can be used to extract the strong coupling constant, to constrain TMD  distribution functions, and to tag  boosted objects produced in high energy collisions. 
  It can be used to extract the strong coupling constant and to constrain TMD  distribution functions. Closely related energy-correlation shape variables have also been used to tag  boosted objects produced in high energy collisions. The opportunities to study EEC at the future Electron-Ion Collider are also highlighted.    
\end{abstract}

\maketitle

\tableofcontents
\newpage

\section{Introduction}

Event shape observables (such as thrust, C-parameter, etc.) are measures of  the energy flow,  multiple particle correlations, and the radiative patterns in high energy collisions. They have been extensively investigated at various colliders and, over the past several decades, have played a central role in our understanding of the perturbative and  non-perturbative aspects of Quantum Chromodynamics (QCD). 

The energy-energy correlator (EEC) is such an observable, which was originally introduced in the context of $e^+e^-$ collisions as an alternative to the thrust family of event shapes. The EEC is defined as follows~\cite{Basham:1978bw}, 
\begin{align}
    \text{EEC}(\chi) =  \sum_{a,b} \int \frac{d\sigma_{e e\to a+b+X}}{\sigma}\, w_{ab}\, \delta(\cos\chi_{ab}-\cos\chi) \, , 
\end{align}
where the sum runs over all the hadron pairs $(a,b)$ and the cross section is weighted by $w_{ab} \equiv 2E_a E_b /s$, which is the product of the energies of $a$ and $b$ normalized by the center-of-mass energy of the system. EEC measures the energy correlations as a function of the opening angle $\chi_{ab}$ between particles $a$ and $b$. 
In the literature the EEC is often written in terms of 
\begin{align} \label{eq:z}
 z \equiv \frac{1}{2} (1 - \cos\chi) = \sin^2\left(\frac{\chi}{2}\right)
\,,\end{align}
which allows to work with a variable $ z \in [0,1]$. Significant theoretical effort has been devoted to understanding this observable, including fixed-order calculations~\cite{Dixon:2018qgp,Luo:2019nig,Belitsky:2013xxa,Belitsky:2013bja,Belitsky:2013ofa,Henn:2019gkr}, and QCD factorization and resummation in the back-to-back $z\to 1$ ($\chi \to \pi$)  and collinear $z\to 0$ ($\chi \to 0$) limits~\cite{Moult:2018jzp,Dixon:2019uzg,Kologlu:2019mfz,Korchemsky:2019nzm,Chen:2020uvt} where logarithmic singularities spoil the convergence of the perturbative series and therefore resummation techniques are necessary to get reliable predictions in these regions. It is important to note that the two limits have very different behaviour as the small angle limit is characterized by a single logarithmic series, while in the back-to-back limit there is double logarithmic behaviour generating a Sudakov-like peak.  Very recently~\cite{Ebert:2020sfi}, the back-to-back limit was calculated at N$^3$LL$^\prime$ accuracy after obtaining the $\mathcal{O}(\alpha_s^3)$ singular distributions constituting the first example of an event shape resummed analytically at this level of accuracy.

%%%%%%%%%%%%%%%%%
\begin{figure}
 \centering
 \includegraphics[width=0.7\textwidth]{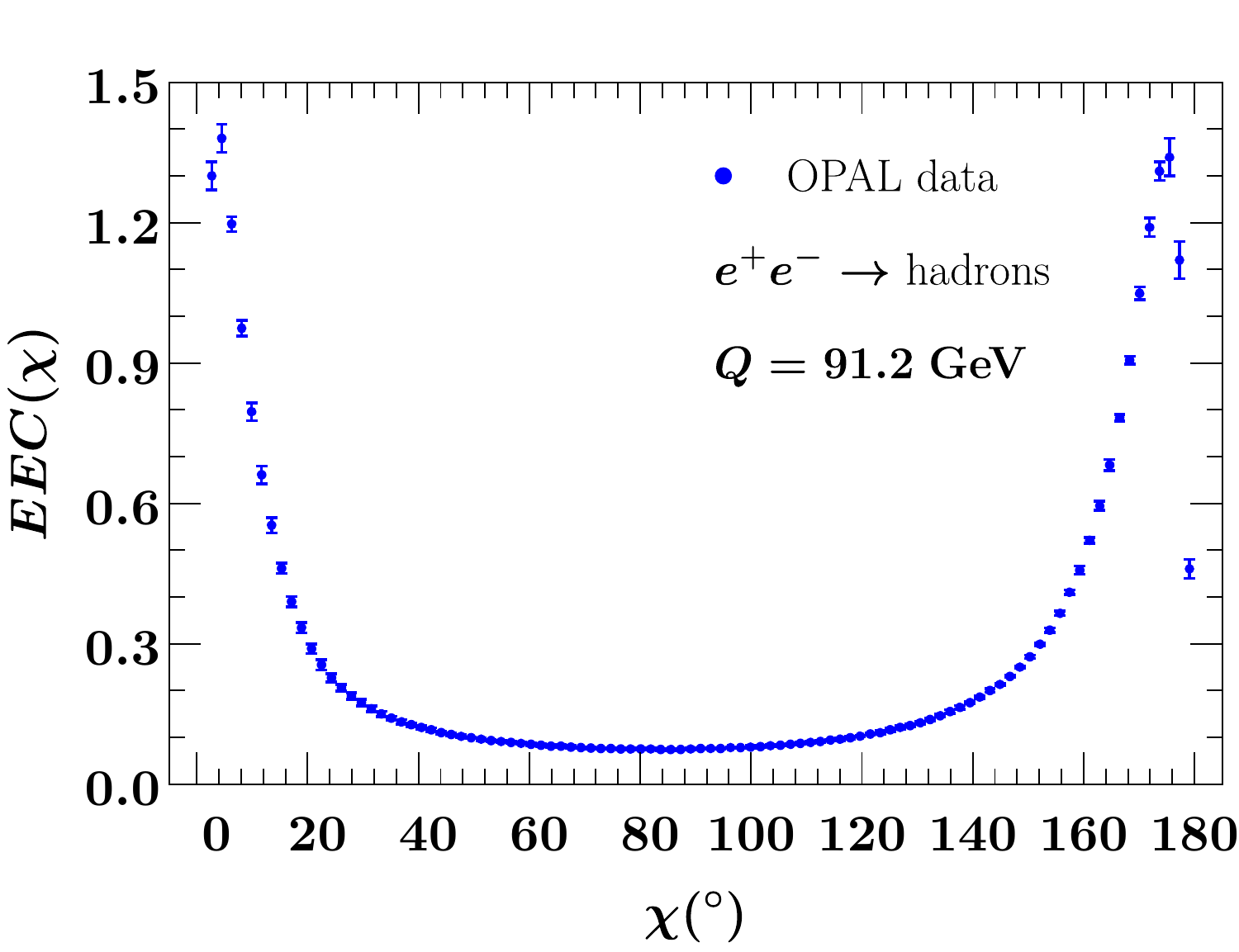}
 \caption[tt]{Experimental measurement of the EEC at LEP. The figure uses experimental data from the OPAL collaboration tabulated in \cite{OPAL:1993pnw}.  The presence of a peaked structure in the back-to-back ($\chi \to \pi$)  and collinear ($\chi \to 0$) limits is evident. Analytically, these enhancements manifest themselves as logarithmic singularities in the perturbative expansion, therefore requiring an all-order treatment of such contributions via resummation techniques.}
 \label{fig:EEC_data}
\end{figure}
%%%%%%%%%%%%%%%%%

At hadronic colliders, an adaptation of EEC known as transverse-energy-energy correlation (TEEC) considers only the momenta in the transverse plane in order to construct the corresponding observable~\cite{Ali:1984yp}. TEEC was calculated including  next-to-leading order (NLO) QCD corrections in Ref.~\cite{Ali:2012rn}, and NNLL resummation in the dijet limit was accomplished in Ref.~\cite{Gao:2019ojf}.

For observables like the EEC and TEEC  soft radiation contributes only through recoil to the energetic collinear particles, since direct contributions from soft emissions are suppressed by the energy weighting factor.  Thus, we should anticipate to have smaller non-perturbative corrections compared to other event shape variables. Moreover, owing to the high perturbative accuracy achieved both in resummed and fixed order calculations~\cite{Moult:2018jzp,Gao:2019ojf,Li:2020bub,Ebert:2020sfi}, complemented  by high precision measurements~\cite{Akrawy:1990hy,Decamp:1990nf,Adeva:1991vw,Abe:1994wv,ATLAS:2015yaa,Aaboud:2017fml,ATLAS:2020mee}, EEC and TEEC observables offer an opportunity for precision studies in QCD.  In particular, EEC and TEEC have been used for precise extractions of the strong coupling constant, for a recent review see Section 9 of Ref.~\cite{10.1093/ptep/ptaa104}.

\section{The EEC in the Back-to-Back Limit}
For $z\to 1$, the energy-energy correlation presents a double logarithmic series at each order in perturbation theory. 
As $z$ gets closer and closer to 1, i.e. as we look at the energy correlations of particles that are closer and closer to a back-to-back dijet configuration, these large logarithmic contributions spoil the perturbative convergence of the fixed order calculation.
Therefore, in order to get reliable predictions it is then necessary to have an understanding of this asymptotics to all orders in the coupling.
\subsection{Factorization and Connection to TMD Fragmentation Functions}
We can understand the all-order behaviour of the EEC in the back-to-back limit by using Soft and Collinear Effective Theory (SCET) \cite{Bauer:2000ew, Bauer:2000yr, Bauer:2001ct, Bauer:2001yt}.
In this asymptotic the EEC can be seen as a weighted cross section of the small-$q_T$ distribution for di-hadron production in $e^+ e^-$ annihilation ~\cite{Collins:1981uk, Collins:1981va, Moult:2018jzp,Ebert:2020sfi},
\begin{align} \label{eq:EEC_qT_fact}
 \lim_{z\to1} \frac{\df\sigma}{\df z}
 = \int_0^1 \df z_1 \df z_2 \, \frac{z_1 z_2}{2} \int\df^2\qt \, \delta\biggl(1 - z - \frac{q_T^2}{Q^2} \biggr)
   \lim_{q_T\to0} \sum_{h_1, h_2} \frac{\df\sigma_{e^+ e^- \to h_1 h_2}}{\df z_1 \df z_2 \df^2\qt}
\,,\end{align}
where $z_{1,2} = (P_{1,2} \cdot q)/ q^2$ is the longitudinal momenta of the hadrons and $q^\mu = p_{e^+}^\mu + p_{e^-}^\mu$ is the momentum of the color singlet source.
Importantly the $\qt$ and the EEC measurements become related at small $\qt$ since 
\beq
	1 - z  =\cos^2\left(\frac{\chi}{2}\right) = \frac{q_T^2}{Q^2}\left[1 + \cO\left(\frac{q_T^2}{Q^2}\right) \right] \, .
\eeq
At small $\qt$, the differential di-hadron production cross section in $e^+ e^-$ annihilation, $\frac{\df\sigma_{e^+ e^- \to h_1 h_2}}{\df z_1 \df z_2 \df^2\qt}$, factorizes as~\cite{Collins:1981uk, Collins:1981va}
\begin{align} \label{eq:qT_fact_q}
 \frac{\df\sigma_{e^+ e^- \to h_1 h_2}}{\df z_1 \df z_2 \df^2\qt} &
 = \born H_{q\bar q}(Q, \mu)
   \int\!\frac{\df^2\bt}{(2\pi)^2} \, e^{\img \qt \cdot \bt}
   \tilde D_{h_1 / q}\Bigl(z_1, b_T, \mu, \frac{\nu}{Q}\Bigr) \tilde D_{h_2 / \bar q} \Bigl(z_2, {b_T}, \mu, \frac{\nu}{Q}\Bigr)
   \nn\\&\quad \times
   \tilde S_q(b_T, \mu, \nu)
   ~\times~\Bigl[1 + \cO\Bigl(\frac{q_T^2}{Q^2}\Bigr)\Bigr]
\,,\end{align}
where $\tilde D_{h_1 / q}\Bigl(z_1, b_T, \mu, \frac{\nu}{Q}\Bigr)$  is a Transverse Momentum Dependent Fragmentation Function (TMDFF) and it is convenient to express the result in impact parameter space, with $\bt$ being the conjugate variable to $\qt$, as the functions factorize as simple products in this space rather than as convolutions. 
Note that for perturbative values of $\qt$, i.e. $\qt \gg \lqcd$, or equivalently for $b_T \lesssim \lqcd^{-1}$, the TMDFFs obey an operator product expansion onto longitudinal (collinear) fragmentation functions $d_{h / q}$ via
\begin{align} \label{eq:F_matching}
 \tilde D_{h / q}\Bigl(z, b_T, \mu, \frac{\nu}{Q}\Bigr)
 = \sum_{q'} \int_z^1 \frac{\df z'}{z'} d_{h/q'}\Bigl(\frac{z}{z'}\Bigr)
   \tilde \cK_{q q'}\Bigl(z', b_T, \mu, \frac{\nu}{Q}\Bigr)
   + \cO(b_T^2 \lqcd^2)
\,,\end{align}
where here $\tilde \cK_{q q'}$ is a matching kernel that can be calculated in perturbation theory.

\begin{figure}[t]
 \centering
 \includegraphics[width=0.3\textwidth]{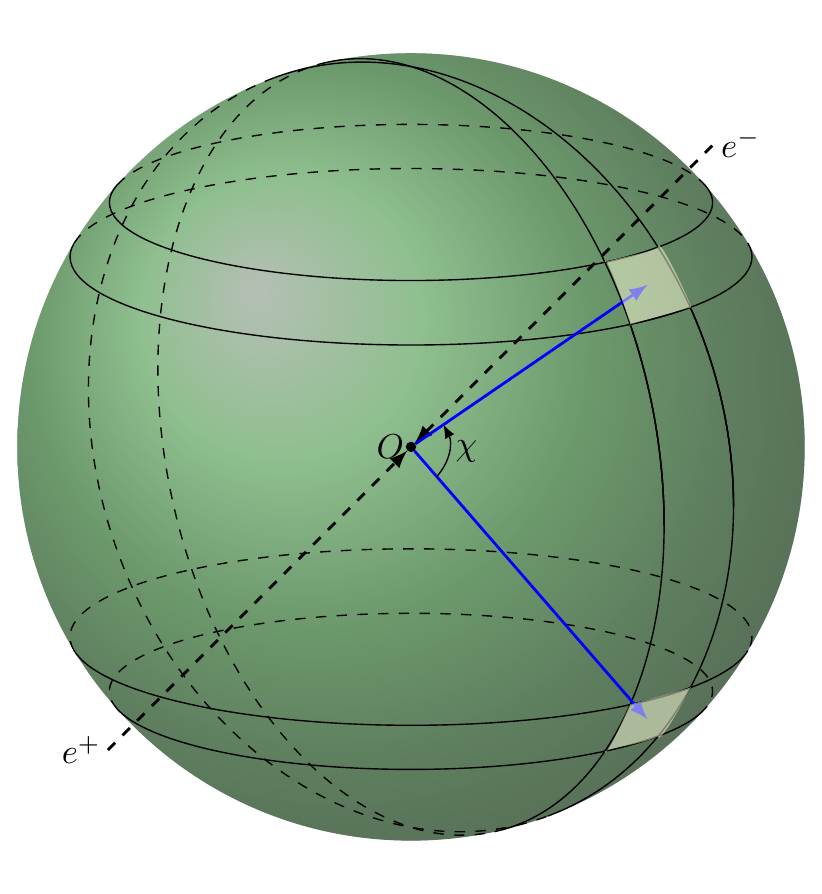}
 \caption{Graphical representation of the Energy Energy Correlator in electron positron annihilation. The EEC measures the energy flowing through two calorimeters at infinity separated by a given angle $\chi$. Figure from \refcite{Moult:2018jzp}.}
 \label{fig:EECzto1_picture}
\end{figure}

Using \eqs{EEC_qT_fact}{qT_fact_q} we obtain the factorization theorem for the Energy Energy Correlation in the back-to-back limit \cite{Moult:2018jzp}
\begin{align} \label{eq:EEC_fact_thm_q}
 \frac{\df\sigma}{\df z} &
 = \frac{\born}{2} H_{q\bar q}(Q, \mu)
   \int\frac{\df^2\bt \, \df^2\qt}{(2\pi)^2} e^{\img \qt \cdot \bt} \delta\biggl(1 - z - \frac{q_T^2}{Q^2} \biggr)
   J_q\Bigl(b_T, \mu, \frac{\nu}{Q}\Bigr) J_{\bar{q}}\Bigl(b_T, \mu, \frac{\nu}{Q}\Bigr) \tilde S_q(b_T, \mu, \nu)
   \nn\\&\quad \times [ 1 + \cO(1-z) ]
%%%
\nn\\&
%%%
 = \frac{\born}{8} H_{q\bar q}(Q,\mu) \int_0^\infty \df (b_T Q)^2 \, J_0\bigl(b_T Q \sqrt{1-z}\bigr)
   J_q\Bigl(b_T, \mu, \frac{\nu}{Q}\Bigr) J_{\bar{q}}\Bigl(b_T, \mu, \frac{\nu}{Q}\Bigr) \tilde S_q(b_T, \mu, \nu)
   \nn\\&\quad \times [ 1 + \cO(1-z) ]
\,.\end{align}
Here $J_0$ is the 0-th order Bessel function, which naturally arises from the Fourier Transform with radial symmetry. 
$ H_{q\bar q}(Q, \mu)$ is the hard function for $e^+ e^-$ annihilation to quark-antiquark and it is related to the IR finite part of the form factor, which has now been calculated at N$^4$LO in \refcite{Lee:2022nhh} and was known up to N$^3$LO for a long time \cite{Kramer:1986sg, Matsuura:1987wt, Matsuura:1988sm, Gehrmann:2005pd, Moch:2005tm, Moch:2005id, Baikov:2009bg, Lee:2010cga, Gehrmann:2010ue}. We normalize the hard function by the born cross section $\born$ such that $H = 1 + \cO(\alpha_s)$.
$\tilde S_q(b_T, \mu, \nu)$ is the EEC soft function, which is equivalent to the $q_T$ soft function normally appearing in transverse momentum distributions at small $q_T$ for  color singlet production at the LHC, see \refcite{Moult:2018jzp} for details on this relation. 
The $q_T$ soft function has been obtained at 3 loops via bootstrapping techniques in \refcite{Li:2016ctv} and later confirmed by direct calculation in \refscite{Luo:2019szz,Ebert:2020yqt}.
$J_q\Bigl(b_T, \mu, \frac{\nu}{Q}\Bigr)$ is the EEC jet function. It describes the dynamics of energetic collinear particles under the EEC measurement constraint. Because of \eq{EEC_qT_fact}, there is a key relation between the EEC jet function and the TMD Fragmentation Functions
\begin{align} \label{eq:def_Jq}
 J_q\Bigl(b_T, \mu, \frac{\nu}{Q}\Bigr) &=
 \sum_{h} \int_0^1 \df z \, z  \, \tilde D_{h / q}\Bigl(z, b_T, \mu, \frac{\nu}{Q}\Bigr) 
 \nn \\&= 
 \sum_{q'} \int_0^1 \df z' \, z' \, \tilde \cK_{q q'}\Bigl(z', b_T, \mu, \frac{\nu}{Q}\Bigr)
\,,\end{align}
where the last line holds because of the sum rule for the fragmentation function,
\begin{align} \label{eq:sum_rule}
 \sum_h \int_0^1 \df z \, z \, d_{h / q}(z,\mu) = 1
\,.\end{align}
The EEC jet function has been calculated to N$^3$LO in \refcite{Ebert:2020sfi}. This result has been obtained in conjunction to the calculation at N$^3$LO for the TMD Fragmentation Functions matching kernels~\cite{Ebert:2020qef} via an analytic continuation of the TMDPDFs of \refcite{Ebert:2020yqt} using the framework of collinear expansion for color singlet cross section \cite{Ebert:2020lxs}. Note that the TMDFF matching kernels where also calculated independently in \refcite{Luo:2020epw} using different techniques \cite{Luo:2019szz,Chen:2020uvt}.

\subsection{Resummation}

{
\renewcommand{\arraystretch}{1.2}
\begin{table}[pt]
\centering
 \begin{tabular}{l|c|c|c|c} \hline\hline
  Accuracy & $H$, $J$, $S$ & $\GammaC(\as)$ & $\gamma_i(\as)$ & $\beta(\as)$ \\\hline
  LL           & Tree level & $1$-loop & --       & $1$-loop \\\hline
  NLL          & Tree level & $2$-loop & $1$-loop & $2$-loop \\\hline
  NLL$^\prime$ & $1$-loop   & $2$-loop & $1$-loop & $2$-loop \\\hline
  NNLL         & $1$-loop   & $3$-loop & $2$-loop & $3$-loop \\\hline
  NNLL$^\prime$& $2$-loop   & $3$-loop & $2$-loop & $3$-loop \\\hline
  N$^3$LL         & $2$-loop   & $4$-loop & $3$-loop & $4$-loop \\\hline
  N$^3$LL$^\prime$& $3$-loop   & $4$-loop & $3$-loop & $4$-loop \\\hline
  N$^4$LL         & $3$-loop   & $5$-loop & $4$-loop & $5$-loop \\\hline
  N$^4$LL$^\prime$& $4$-loop   & $5$-loop & $4$-loop & $5$-loop \\\hline
 \hline
 \end{tabular}
\caption{%
Resummation accuracy in terms of the perturbative order of boundary terms, anomalous dimensions and beta function. Here $\gamma_i$ is a short hand for all the non-cusp anomalous dimensions, including the rapidity anomalous dimension. See \Eq{RGEs} for details.
}
\label{tbl:log_counting}
\end{table}
\renewcommand{\arraystretch}{1.0}
}

The Hard, Jet and Soft functions appearing in \eq{EEC_fact_thm_q} contain logarithms of the ratio between the characteristic scales of the hard, collinear, and soft radiation and of the $\mu$ and $\nu$ renormalization scales.
These logarithms naturally arise in the fixed order calculation of these objects after regulating both UV and rapidity divergences. Using the standard QFT technique of removing these divergences with appropriate counterterms and imposing the independence of the bare quantities on the $\mu$ and $\nu$ renormalization scales, one derives the Renormalization Group Equations (RGEs) for the Hard, Jet and Soft functions.
The $\mu$-evolution is dictated by the following RGEs
\begin{alignat}{3} \label{eq:RGEs}
 \frac{\df}{\df \ln \mu} \ln H_i(Q,\mu) &
 = \gamma_H^i(Q,\mu)
\,,\nn\\
 \frac{\df}{\df \ln \mu} \ln J_i(b_T,\mu,\nu/Q) &
 = \tilde\gamma_J^i(\mu,\nu/Q)
\,,\nn\\
 \frac{\df}{\df \ln \mu} \ln \tilde S_i(b_T,\mu,\nu) &
 = \tilde\gamma_S^i(\mu,\nu)
\,.\end{alignat}
The anomalous dimensions can be organized in the sum of two terms.
The first one drives the leading double logarithmic behaviour and it is proportional to the cusp anomalous dimension \cite{Korchemsky:1987wg}, while the second one is the non-cusp piece and it determines the logarithms beyond leading logarithmic accuracy 
\begin{align} \label{eq:mu_anom_dims}
 \gamma_H^i(Q,\mu) &= 4 \GammaC^i[\as(\mu)] \ln\frac{Q}{\mu} + 4 \gamma_i[\as(\mu)]
\,,\nn\\
 \tilde\gamma_J^i(\mu,\nu/Q) &= 2 \GammaC^i[\as(\mu)] \ln\frac{\nu}{Q} + \tilde\gamma_J^i[\as(\mu)]
\,,\nn\\
 \tilde\gamma_S^i(\mu,\nu) &= 4 \GammaC^i[\as(\mu)] \ln\frac{\mu}{\nu} + \tilde\gamma_S^i[\as(\mu)]
\,.\end{align}
Note that the cusp is known to 4 loops \cite{Henn:2019swt} and the non-cusp hard anomalous dimension $\gamma_i$ is related to the collinear anomalous dimension which has also been obtained at 4 loops in QCD \cite{ vonManteuffel:2020vjv}. The jet and soft non cusp anomalous dimensions are related to the threshold anomalous dimension and to the DGLAP kernel at threshold and are therefore know to 3 loops.
Because of the presence of rapidity divergences in TMD observables, such as the EEC in this limit, the Jet and Soft function obey a Rapidity Renormalization Group Equation \cite{Chiu:2012ir}
\begin{alignat}{3} \label{eq:RRGE}
 \frac{\df}{\df \ln \nu} \ln J_i(b_T,\mu,\nu/Q) &= -\frac12 \tilde\gamma_\nu^i(b_T,\mu)
\,,\nn\\
 \frac{\df}{\df \ln \nu} \ln \tilde S_i(b_T,\mu,\nu) &= \tilde\gamma_\nu^i(b_T,\mu)
\,,\end{alignat}
where $\tilde\gamma_\nu^i(b_T,\mu)$ is the rapidity anomalous dimension that has been calculated to N$^3$LO in \refcite{Li:2016ctv}.
We can solve analytically the RGE equations \eqs{RGEs}{RRGE}, fixing the boundary at the scale $\mu_H$ for the Hard function, $\mu_J$ and $\nu_J$ for the jet function, and $\mu_S$ and $\nu_S$ for the soft function.
Using the solution of the RGEs we obtain the analytic formula for the resummation of the EEC in the back-to-back limit \cite{Moult:2018jzp,Ebert:2020sfi}  
\begin{align} \label{eq:EEC_resummed}
 \frac{\df\sigma}{\df z} &
 = \frac{\born}{8} \int_0^\infty \!\!\df (b_T Q)^2 \, J_0\bigl(b_T Q \sqrt{1-z}\bigr)
     H_{q\bar{q}}(Q,\mu_H) J_q\Bigl(b_T, \mu_J, \frac{\nu_J}{Q}\Bigr) J_{\bar{q}}\Bigl(b_T, \mu_J, \frac{\nu_J}{Q}\Bigr) \tilde S_q(b_T, \mu_S, \nu_S)
 \nn\\&\quad \times
 \exp\left[ \int_{\mu_H}^\mu \frac{\df\mu'}{\mu'} \gamma_H^q(Q, \mu')
          + 2 \int_{\mu_J}^\mu \frac{\df\mu'}{\mu'} \tilde\gamma_J^q(\mu', \nu_J/Q)
          + \int_{\mu_S}^\mu \frac{\df\mu'}{\mu'} \tilde\gamma_S^q(\mu', \nu_S) \right]
 \Bigl(\frac{\nu_J}{\nu_S}\Bigr)^{\tilde\gamma_\nu^q(b_T, \mu)}
 \nn\\&\quad \times
 \bigl[ 1 + \cO(1-z) \bigr]
\,.\end{align}
Note that perturbative power corrections to \eq{EEC_resummed} have been studied and resummed in $\mathcal{N}=4$ SYM in \refcite{Moult:2019vou} and they necessitate the treatment of subleading power rapidity divergenges \cite{Ebert:2018gsn} analogous to TMD distributions for color singlet production at the LHC.
The \emph{canonical} choice for the boundaries is 
\begin{align} \label{eq:canonical_scales}
 \mu_H \sim Q \,,\quad &\mu_J \sim \frac{b_0}{b_T} \,, \quad \mu_S \sim \frac{b_0}{b_T}
\,,\nn\\
 &\nu_J \sim Q \,,\quad~~ \nu_S \sim \frac{b_0}{b_T}
\,,\end{align}
which is the choice of scales that renders all the explicit logarithms in the boundary functions vanishing.
Resummation uncertainties can be obtained by varying the choice of scales around the canonical values, see \refcite{Ebert:2020sfi} for details.

In \Eq{EEC_resummed}, both the boundary functions $H,J,S$ as well as the anomalous dimensions can be calculated in perturbation theory.
The accuracy of the resummed prediction depends on the order at which these different ingredients are calculated. In Table \ref{tbl:log_counting} we provide a summary for the counting of the resummation accuracy as a function of the perturbative order of the anomalous dimensions and boundaries entering \eqs{RGEs}{RRGE}. Note that N$^n$LL$^\prime$ means \emph{higher} accuracy than simple N$^n$LL, as it includes the full fixed order ingredients at $\cO(\alpha^n_s)$.

\subsection{Event Shapes at 3 Loops in QCD and $\alpha_s$ Extraction}

In \refcite{Ebert:2020sfi} the calculation of the EEC quark and gluon jet functions at N$^3$LO determined the last missing ingredient for the complete analytic result at $\cO(\alpha_s^3)$ in QCD for the back-to-back limit of this observable. At this order the $z\to 1$ asymptotic of the EEC reads
{\allowdisplaybreaks
\small
\begin{align} \label{eq:EEC_z1_N3LO_ee}
 \frac{1}{C_F} \frac{\df \bar\sigma^{(3)}}{\df z} &
 = -4 C_F^2 \cL_5(\zb) + \cL_4(\zb) \Bigl[-30 C_F^2 - \frac{220}{9} C_F C_A  + \frac{40}{9} C_F n_f \Bigr]
 \nn\\&\quad
  + \cL_3(\zb) \Bigl[ C_F^2 (-16 \zeta_2-104) +\frac{88}{9} C_F n_f + C_F C_A \Bigl(-16 \zeta_2-\frac{388}{9}\Bigr) - \frac{242}{9} C_A^2 + \frac{88}{9} C_A n_f - \frac{8}{9} n_f^2 \Bigr]
 \nn\\&\quad
  + \cL_2(\zb) \Bigl[
     C_F^2 (-144 \zeta_2-16 \zeta_3-189)
   + C_F C_A \Bigl(-\frac{592}{3}\zeta_2-72 \zeta_3+\frac{244}{3}\Bigr)
   \nn\\&\hspace{2cm}
   + C_F n_f \Bigl(\frac{88}{3}\zeta_2-\frac{40}{3}\Bigr)
   + C_A^2\Bigl(\frac{2471}{27}-\frac{88}{3}\zeta_2 \Bigr)
   + C_A n_f \Bigl(\frac{16}{3}\zeta_2 - \frac{760}{27}\Bigr)
   + \frac{44n_f^2}{27}
  \Bigr]
 \nn\\& \quad
  + \cL_1(\zb) \Bigl[
    C_F^2 \Bigl(-\frac{542}{3} -412 \zeta_2 + 224 \zeta_3-192 \zeta_4\Bigr)
    + C_F C_A \Bigl(-\frac{2900}{9}\zeta_2-\frac{1688}{3}\zeta_3-8 \zeta_4+\frac{3797}{9}\Bigr)
    \nn\\&\hspace{2cm}
    + C_F n_f \Bigl(\frac{536}{9}\zeta_2+\frac{32}{3}\zeta_3-\frac{479}{9}\Bigr)
    + C_A^2 \Bigl(-\frac{916}{9}\zeta_2-44 \zeta_4-\frac{2354}{81}\Bigr)
    \nn\\&\hspace{2cm}
    + C_A n_f\Bigl(\frac{448}{9}\zeta_2 +16 \zeta_3-\frac{380}{81}\Bigr)
    + n_f^2\Bigl(\frac{124}{81}-\frac{16}{3}\zeta_2\Bigr)
   \Bigr]
 \nn\\&\quad
  + \cL_0(\zb) \Bigl[
      C_F^2 \Bigl(64 \zeta_3 \zeta_2-402\zeta_2+332 \zeta_3-552 \zeta_4+48 \zeta_5-\frac{169}{2}\Bigr)
    \nn\\&\hspace{2cm}
    + C_F C_A \Bigl(-128\zeta_3 \zeta_2+\frac{212}{3}\zeta_2 -\frac{2812}{9}\zeta_3 -\frac{1342}{3}\zeta_4 - 120 \zeta_5+\frac{3358}{9}\Bigr)
    \nn\\&\hspace{2cm}
    + C_F n_f \Bigl(-\frac{20}{3}\zeta_2 -\frac{296}{9}\zeta_3 + \frac{244}{3}\zeta_4 - \frac{623}{18}\Bigr)
    + C_A^2 \Bigl(\frac{4420}{9}\zeta_2 -\frac{560}{9}\zeta_3-\frac{326}{3}\zeta_4 - 40 \zeta_5-\frac{4241}{27}\Bigr)
    \nn\\&\hspace{2cm}
    + C_A n_f\Bigl(-\frac{1508}{9} \zeta_2 +\frac{184}{9}\zeta_3 + \frac{56}{3}\zeta_4 + \frac{1414}{27}\Bigr)
    + n_f^2 \Bigl(\frac{112}{9}\zeta_2 + \frac{16}{9}\zeta_3 - \frac{98}{27}\Bigr)
    \Bigr]
 \nn\\&\quad
 + \delta(\zb) \Bigl[ C_F^2 \Bigl(-\frac{337}{3} -\frac{1049}{3}\zeta_2 +\frac{530}{3} \zeta_3 + 512 \zeta_2 \zeta_3 - 64 \zeta_3^2 -1396 \zeta_4+\frac{3136 }{3} \zeta_5-672 \zeta_6\Bigr)
    \nn\\&\hspace{1.5cm}
    + C_F C_A \Bigl(\frac{10169}{27} +\frac{2729}{3}\zeta_2 -\frac{22070}{9}\zeta_3 +\frac{2176 \zeta_4}{9}+ 528 \zeta_5 + 22 \zeta_6 -288 \zeta_2 \zeta_3 + 64 \zeta_3^2\Bigr)
    \nn\\&\hspace{1.5cm}
    + C_F n_f \Bigl(-\frac{148}{27} -\frac{985 \zeta_2}{9}+\frac{3340 \zeta_3}{9}+\frac{58 \zeta_4}{9}-\frac{368 \zeta_5}{3} -\frac{224}{3} \zeta_2 \zeta_3 \Bigr)
    \nn\\&\hspace{1.5cm}
    + C_A^2 \Bigl(-\frac{55504}{81} -\frac{3968}{81}\zeta_2 +\frac{39337}{27}\zeta_3 +\frac{3815}{18}\zeta_4 - \frac{2720}{3}\zeta_5 -\frac{700}{3}\zeta_2 \zeta_3 + 59 \zeta_6  -56 \zeta_3^2 \Bigr)
    \nn\\&\hspace{1.5cm}
    + C_A n_f \Bigl(\frac{15626}{81} - \frac{3326}{81}\zeta_2 - \frac{3788}{27}\zeta_3 - \frac{290}{9}\zeta_4 + 80\zeta_5 + 72 \zeta_2 \zeta_3 \Bigr)
    \nn\\&\hspace{1.5cm}
    + n_f^2 \Bigl(-\frac{1048}{81} + \frac{616}{81}\zeta_2 - \frac{464}{27}\zeta_3 - \frac{16}{9}\zeta_4 \Bigr)
    + N_{F,V} \frac{d_{abc}d^{abc}}{N_r} (2+5 \zeta_2+\frac{7}{3} \zeta_3-\frac{\zeta_4}{2}-\frac{40}{3} \zeta_5)
\Bigr]
\,,\end{align}
}%
where $\zb \equiv 1-z$, $\cL_n(\zb)$ indicates the standard plus distribution, $ \cL_n(\zb) \equiv \left[\frac{\log^n(\zb)}{\zb}\right|_+$, and $N_{F,V}$, following the notation of \refcite{Gehrmann:2010ue}, is the charge of the singlet contribution, where the vector boson that sources the hard scattering event is coupled to a closed quark loop.

\begin{figure}[h]
 \centering
 \includegraphics[width=0.7\textwidth]{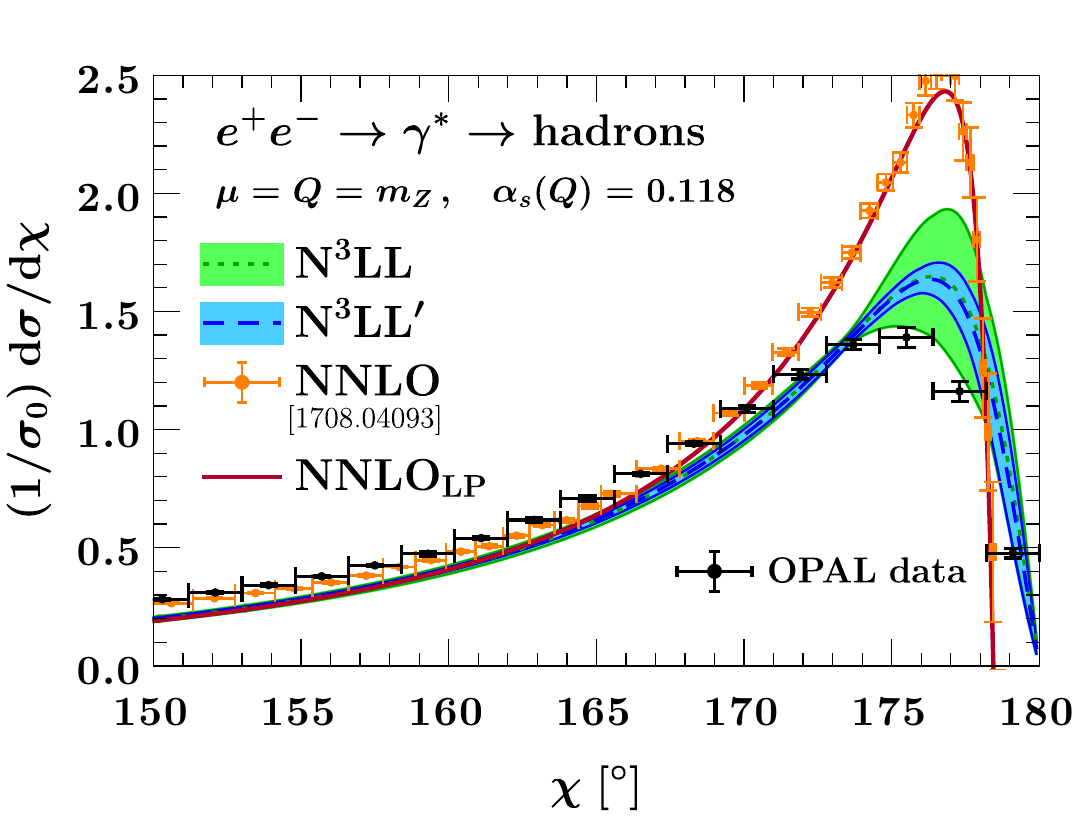}
 \caption[dsadsasd]{EEC partonic spectrum at N$^3$LL and N$^3$LL$^\prime$ resummation accuracy from \refcite{Ebert:2020sfi} compared to the analytic leading power spectrum at $\cO(\alpha_s^3)$, the NNLO numerical  fixed order calculation of~\refcite{Tulipant:2017ybb} and the experimental data from the OPAL collaboration at LEP \cite{OPAL:1993pnw}. Figure from \refcite{Ebert:2020sfi}.}
 \label{fig:EEC_data_comp}
\end{figure}

With \eq{EEC_z1_N3LO_ee} and the N$^3$LO anomalous dimensions, the resummation of the EEC in the back-to-back limit was carried out at N$^3$LL$^\prime$ \cite{Ebert:2020sfi}, constituting the first event shape to be resummed analytically at this order\footnote{Note that for both Thrust and C-Parameter the $\cO(\alpha_s^3)$ spectrum is not known analytically due to the fact that the soft function is not known at this order. However, very good numerical approximation can be made for it allowing the resummation of these event shapes at N$^3$LL$^\prime$ with a small additional uncertainty due to this unknown 3 loop ingredient \cite{Abbate:2010xh,Hoang:2014wka}.}.
In \fig{EEC_data_comp} we show the impact of N$^3$LL and N$^3$LL$^\prime$ resummation on the EEC spectrum in the back to back region compare to the numerical fixed order result from \refcite{Tulipant:2017ybb} and experimental measurements from \fig{EEC_data} of the OPAL collaboration at LEP \cite{OPAL:1993pnw}.
We see that the resummation cures the pathological behaviour of the fixed order cross section for large angles and that the inclusion of N$^3$LL' corrections reduces the perturbative uncertainty to a level that is now comparable to the experimental accuracy.

Thanks to its high sensitivity to QCD radiation the EEC is a natural candidate for the extraction of the strong coupling constant. For a recent review  on the extraction of $\alpha_s$ from EEC measurements see Section 9 of Ref.~\cite{10.1093/ptep/ptaa104}.
Given the very high level of control on the perturbative convergence of this event shapes obained with the inclusion of resummation effects up to N$^3$LL$^\prime$, it is possible to improve the precision of such extractions by including in the analysis the large amount of data in the back-to-back region that were originally excluded due to the poor theoretical control of the observable in this region.
However, although the uncertainties are now comparable, it is evident that there is tension between the partonic prediction and the data.
It is crucial to note that for a full phenomenological analysis it will be necessary to include a variety of effects such as:
\begin{itemize}
    \item Electroweak corrections 
    \item Finite quark mass effects
    \item Singlet contributions to the axial current
    \item Hadronization and non perturbative corrections
\end{itemize}
Among them, getting a better understanding of non perturbative corrections, which that are known to be quite sizable, see for example \refcite{Kardos:2018kqj}, is probably the most challenging task. 
All these effects can easily account for few percent shifts of the result and it is important to calculate and include them for future extractions of the strong coupling constant from the EEC, in case we want to include the large amount of data measured in this region.

\section{TEEC at a  Hadron Collider}

Transverse EEC was originally proposed as a quantitative test of perturbative QCD at high energy hadron collisions~\cite{Ali:1984yp}, generalizing the original EEC~\cite{Basham:1978bw}. The idea is to consider transverse energy weighting $E_T = \sqrt{m_T^2 + p_T^2}$ and azimuthal angle correlation, as is common in a hadron collider environment, see Fig.~\ref{fig:TEEC}. Ignoring hadron mass $m_T$, as is appropriate at high energy collider such as the LHC, it simply reduces to transverse momentum weighting. Explicitly, TEEC is defined as
\begin{align}%\label{eq:TEEC_intro}
\frac{\df \sigma}{\df\phi}=\sum\limits_{a,b} \int \df\sigma_{pp\to a+b+X} \frac{2 E_{T,a} E_{T,b}}{ |\sum_i E_{T,i}|^2 }   \delta(\cos\phi_{ab} - \cos\phi)\,, 
\label{eq:TEEC_def}
\end{align}
where $\phi$ is the difference of azimuthal angles between two detected hadrons, while $\df \sigma_{pp\to a+b+X}$ represents the semi-inclusive cross section for production of two hadrons $a$ and $b$ plus anything $X$. To avoid overwhelmed by forward scattering, one can impose kinematical cuts in \eq{TEEC_def}, such as rapidity cut on the detected hadrons, total transverse energy cut, or cut on jet $p_T$. Fortunately, for the kinematical limit discussed below, TEEC is insensitive to these cuts.

%%%%%%%%%%%%%%%%%%%%%%%%%%%%%%%
\begin{figure}
\includegraphics[width=0.3\linewidth]{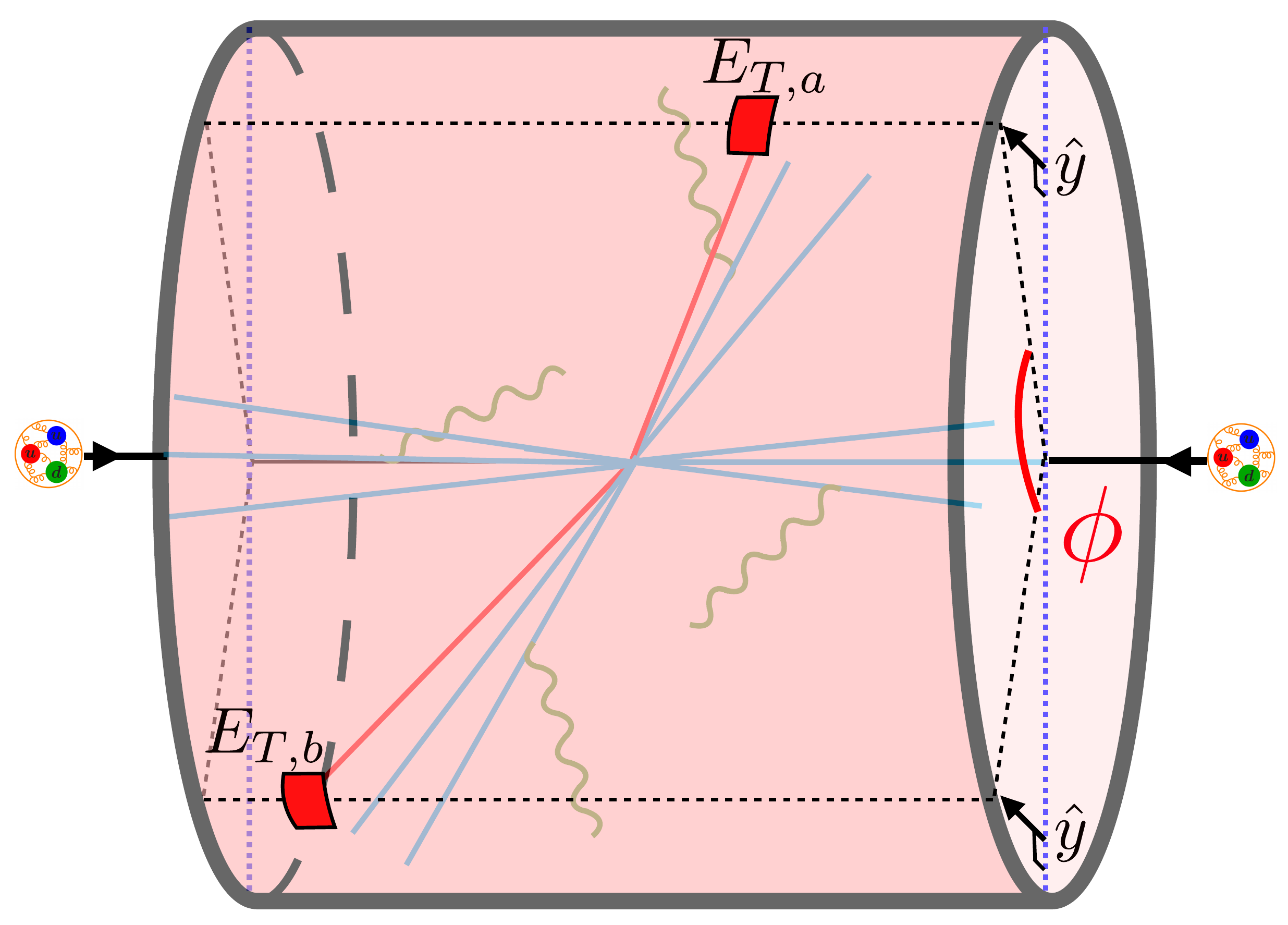}
\caption{A schematic illustration of TEEC at hadron collider. In the $\phi \to \pi$ limit, it measures the momentum in the direction $\hat y$ perpendicular to the scattering plane spanned by the beam and jet axes, outlined in dashed blue. Figure adopted from ref.~\cite{Gao:2019ojf}.
}
\label{fig:TEEC}
\end{figure}
%%%%%%%%%%%%%%%%%%%%%%%%%%%%%%%

A direct analogy of back-to-back limit in EEC translates into back-to-back limit in the azimuthal plane at hadron collider. Following EEC, we can define a scaling variable
\begin{equation}
\tau = \frac{1 + \cos\phi}{2} \,, \qquad \tau \in [0, 1] \,.
\end{equation}
The hadron back-to-back limit corresponds to $\tau \to 0$. Similar to EEC, the back-to-back limit of TEEC is sensitive to Sudakov double logarithms, but with much more complicated kinematics and color evolution. 
Only recently, a resummation formula has been proposed for TEEC in the back-to-back limit~\cite{Gao:2019ojf},
  \begin{align}
\frac{\df\sigma}{\df\tau}
 =&\  \frac{p_T}{16 \pi s^2 (1 + \delta_{f_3 f_4}) \sqrt{\tau}}\sum\limits_{\text{channels}} \frac{1}{N_{\text{init}}}\int \frac{\df y_3 \df y_4 \df p_T^2}{\xi_1\xi_2} \int_{-\infty}^{\infty}\frac{\df b}{2\pi}e^{-2ib\sqrt{\tau} p_T} \mathrm{tr}\big[\mathbf{H}^{f_1 f_2 \to f_3 f_4}(p_T,y^*,\mu) \mathbf{S}(b, y^*, \mu,\nu) \big]\nn \\
&\ \cdot   B_{f_1/N_1}(b,\,\xi_1,\,\mu,\,\nu)\,B_{f_2/N_2}(b,\,\xi_2,\,\mu,\,\nu) J_{f_3}\left(b,\mu,\nu\right)
  J_{f_4}\left(b,\mu,\nu\right) + \text{subleading power} \,.
\label{eq:master}
\end{align}
The sum is over different $2\to 2$ partonic scattering channels $f_1(p_1) f_2(p_2) \to f_3(p_3) f_4(p_4)$, where $N_{\text{init}}$ is the corresponding spin- and color-averaged factor for each channel, $\sqrt{s}$ is the center-of-mass energy,  $y_3$, $y_4$, and $p_T$ are the rapidity and transverse momentum of the two leading partonic jets at the lowest order in perturbation theory, and $\xi_1 = p_T(e^{y_3} + e^{y_4})/\sqrt{s}$ and $\xi_2 = p_T(e^{-y_3} + e^{-y_4})/\sqrt{s}$ are the born-level initial-state momentum fractions. The factorization formula \eq{master} depends on a renormalization/factorization scale $\mu$ and a rapidity scale $\nu$, as is common to TMD-like problem. The beam function $B_{f/N}$ and jet function $J_f$ are simply the one-dimensional projection of TMD beam and fragmentation functions. The novel part which departures most from back-to-back EEC in $e^+e^-$ and DIS is the presence of matrix-valued hard function and soft function in color space. For example, the soft function in Feynman gauge is defined as vacuum expectation value of product of semi-infinite soft Wilson lines, 
\begin{align}
  \label{eq:soft}
\hspace{-0.25cm}\mathbf{S}(b,y^*) = \langle 0 |T[\boldsymbol{O}_{n_1 n_2 n_3 n_4}(0^\mu)] \overline{T}[\boldsymbol{O}_{n_1 n_2 n_3 n_4}^\dagger (b^\mu)] | 0 \rangle \,.
\end{align}
Here $\boldsymbol{O}_{n_1 n_2 n_3 n_4}(x) = \boldsymbol{Y}_{n_1} \boldsymbol{Y}_{n_2} \boldsymbol{Y}_{n_3} \boldsymbol{Y}_{n_4}(x)$, with $\boldsymbol{Y}_{n_i}(x) = \exp[ i \int \df s\, n_i \cdot A(s n_i + x) \mathbf{T}_i]$ a semi-infinite light-like soft Wilson line, and $n_i^\mu = p_i^\mu/p_i^0$ the light-like direction of the incoming or outgoing parton in the partonic center-of-mass frame. The soft function and hard function obey similar matrix-valued evolution equation. For the hard function, the evolution equation control the singularity of on-shell scattering amplitudes, and have received substantial attention recently~\cite{Almelid:2015jia,Almelid:2017qju} due to non-trivial large-angle soft gluon correlation. TEEC thus provides potential collider probes of these correlation effects. 

\begin{figure}[h]
  \centering
  \subfloat[]{%
  \includegraphics[width=0.5\linewidth]{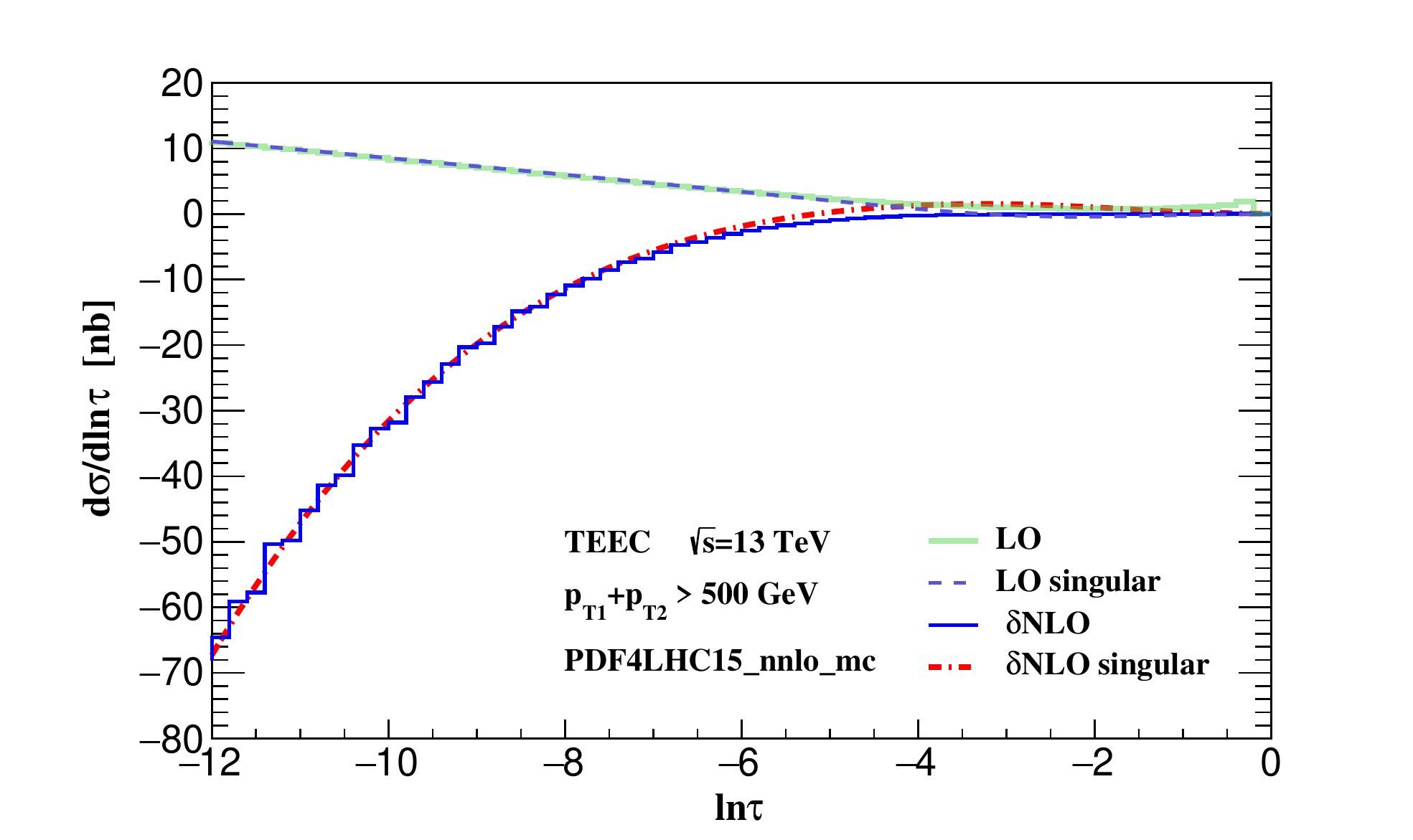}
  \label{fig:asy}
  }
  \subfloat[]{%
    \includegraphics[width=0.5\linewidth]{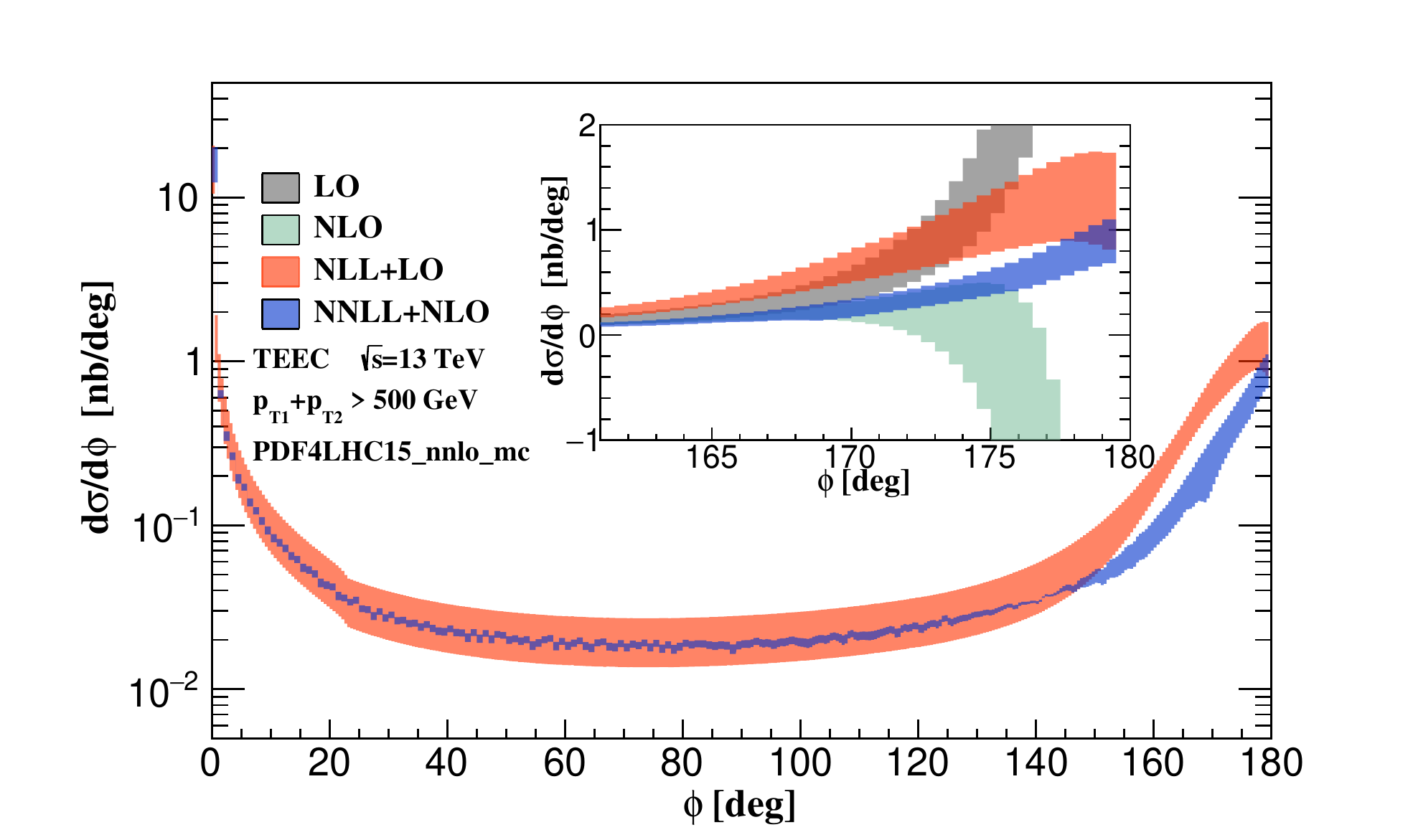}
  \label{fig:resummed}
  }
    \caption{Factorization v.s. fixed order perturbation theory in \fig{asy}. The resummed TEEC distribution matched to fixed order at both NLL+LO and NNLL+NLO in \fig{resummed}. Both figures are adopted from \cite{Gao:2019ojf}.}
\end{figure}

The factorization formula \eq{master} has been validated to high precision against fixed-order prediction from NLOJET++~\cite{Nagy:2001fj,Nagy:2003tz}, as is shown in Fig.~\ref{fig:asy}, adopted from \cite{Gao:2019ojf}. The asymptotic small $\tau$ region agrees nicely between factorization SCET prediction and fixed-order perturbation theory. Fig.~\ref{fig:resummed} shows the NNLL+NLO resummed prediction for TEEC at the LHC from \cite{Gao:2019ojf}. In the back-to-back region~($\phi \to \pi$), fixed-order perturbation theory breaks down and resummation is necessary. It's interesting to note that large corrections are observed when going from NLL+LO to NNLL+NLO. This emphasizes the importance of including higher order corrections. 

%%%%%%%%%%%%%%%%%%%%%%%%%%%%%%%
%%%%%%%%%%%%%%%%%%%%%%%%%%%%%%%

Analytic resummation formula for dijet event shape is rare, while TEEC provides an example where even NNLL accuracy is possible to achieve, thanks to the intrinsic simplicity of the observable. It is therefore important to explore its application to precision QCD physics. We comment on a number of interesting directions below. 

A variant of TEEC has been measured by ATLAS~\cite{ATLAS:2015yaa,ATLAS:2017qir,ATLAS:2020mee}, where correlation of hadrons has been replaced by correlation of jets. The measured has been compared with NLO prediction~\cite{Ali:2012rn} based on NLOJET++~\cite{Nagy:2001fj,Nagy:2003tz}. A determination of $\alpha_S$ has been made using the latest $13$ TeV data~\cite{ATLAS:2020mee}:
\begin{equation}
\alpha_S(m_Z) = 0.1196 \pm 0.0004 (\text{exp.}) ^{+0.0072}_{-0.0105} (\text{theo.}) \,.
\end{equation}
In this determination only data for $|\cos\phi|<0.92$ is used. It would be very interesting to extend the fit down to $\cos\phi < -0.92$, where the cross section is much larger. For that one will need to extend the factorization formula in \eq{master} to incorporate the effects of jet correlation~\cite{Sun:2014gfa}. 
On the other hand, it would be useful if the correlation can be measured directly in terms of hadrons, as NNLL+NLO prediction is directly available. It is well-known that direct measurement of hadron energy correlation leads to large hadronization corrections. However, recent studies shows that hadronization corrections can be substantially reduced if ratio of EEC are measured~\cite{Chen:2020vvp,Komiske:2022enw}. 
It would be interesting to whether similar reduction in hadronization corrections can be achieved in TEEC by forming ratio. It would also be interesting to generalize TEEC to processes of V plus jet production, where V stands for electroweak boson. This process is simpler than dijet production and can achieve even higher accuracy. For the case of jet-boson correlation, this has been done in \cite{Chien:2019gyf,Chien:2020hzh}. It is anticipated that the factorization would be even simpler for hadron-boson correlation. 

At the same time, it would be interesting to push the theory accuracy for TEEC to N$^3$LL and beyond. As originally emphasized in \cite{Gao:2019ojf}, a key motivation for TEEC is its simplicity in factorization, thus allowing probing effects which are deep in the perturbation series, such as factorization breaking effects.  Many of the ingredients for N$^3$LL resummation has become available in recent years, such as three-loop $2 \to 2$ amplitudes~\cite{Caola:2021rqz,Caola:2021izf},  three-loop rapidity anomalous dimension~\cite{Li:2016ctv,Vladimirov:2016dll}, and closely related three-loop TMD beam functions~\cite{Luo:2019szz,Ebert:2020yqt} and fragmentation functions~\cite{Luo:2020epw,Ebert:2020qef}. An important missing ingredient is the three-loop soft function for TEEC at hadron collider, whose calculation may also shed light on the structure of factorization violation.

The factorization formula in \eq{master} can be generalized to include gluon spin correlation effects. At NNLL accuracy such effects are absent for dijet production, but is presented for V+jet production~\cite{Chien:2020hzh}. Starting from N$^3$LL, spin correlation will also be presented in dijet production. In that case the beam function and jet function need to be generalized to incorporate spin effects. The correlation is complicated as it involve both initial state and final state through the hard function. It would be interesting to study to what extent such spin correlation can be measured.  

Finally, it is important to note that TEEC can also be generalized to incorporated track.  Measurement performed on track can lead to smaller systematic uncertainties on experiment side and is therefore very welcome. On theory side, recent studies has shown that the track function formalism~\cite{Chang:2013rca,Chang:2013iba} becomes much simpler when applying to energy correlators~\cite{Chen:2020vvp}. In the collinear limit this has been exploited to obtain moments of anomalous dimension of track function to two loops~\cite{Li:2021zcf,Jaarsma:2022kdd}.  It would be very interesting to apply these results to TEEC.

\section{EEC For Organizing Hard Scattering Processes}
Up to now, we have considered EEC's as a means to probe the distribution of energy correlations surrounding the hard scattering event. The specific hard scattering process is not the main object of study in these cases, and one can sum inclusively over many different hard scattering configurations that all contribute to the same energy-correlation functions. Often though, one wishes to select for specific hard scattering structures, and \emph{distributions} of energy-energy correlations can usefully organize the hard scattering phase space. By putting constraints on the distribution of the sum of over energy correlations in the event or jet, one can select for specific hard scattering geometries. This gives an infra-red and collinear safe means of studying the corresponding S-matrix element (squared), up to a set of cut-offs on phase-space specified by the constraints on the energy-correlation sums. Phenomenologically, this is most useful in tagging boosted decays at hadron-colliders, see Ref.~\cite{Larkoski:2017jix} for review.

In particular, as demonstrated in Refs. \cite{Larkoski:2014gra,Larkoski:2015zka,Larkoski:2015kga,Larkoski:2017cqq}, building on a proposal in Ref.~\cite{Larkoski:2013eya}, by constraining the sums over higher point correlations, we can form shape variables. If we consider the case of a jet with momentum $p_J=Q\frac{n}{2}$, with $n$ a null vector and $\bar{n}$ the conjugate null vector ($n\cdot \bar{n}=2$), these observables take the form:
\begin{align}
    z_i&=\frac{\bar{n}\cdot p_{i}}{Q}\\
    e_{2}^{(\beta)}&=\sum_{i,j\in J} z_i z_j \Big(\frac{2p_i\cdot p_j}{Q^2 z_i z_j}\Big)^{\beta/2},\\
    e_{3}^{(\beta)}&=\sum_{i,j,k\in J} z_i z_j z_k \Big(\frac{2p_i\cdot p_j}{Q^2  z_i z_j}\frac{2p_j\cdot p_k}{Q^2 z_j z_k}\frac{2p_k\cdot p_i}{Q^2 z_k z_i}\Big)^{\beta/2}.
\end{align}
The sums over $i,j,k$, etc., range over all particles in the jet, and $z_i$ is the momentum fraction of the particle in the jet. Higher-point generalizations are straightforward to construct. In the case of investigating the substructure of a jet, one sums only over the particles contained within the jet and replaces the center of mass energy $Q$ with $E_J$, the jet energy, and the energies of the particles by their appropriate projected momentum along the jet direction. As a shape variable, it is easy to see that $e_{2}^{(2)}$ for instance is simply the invariant mass of the jet, and specifying that $e_{2}^{(2)}\ll 1$ selects out a narrow, pencil-like jet. 

\begin{figure}
\begin{center}
\includegraphics[width = 5cm]{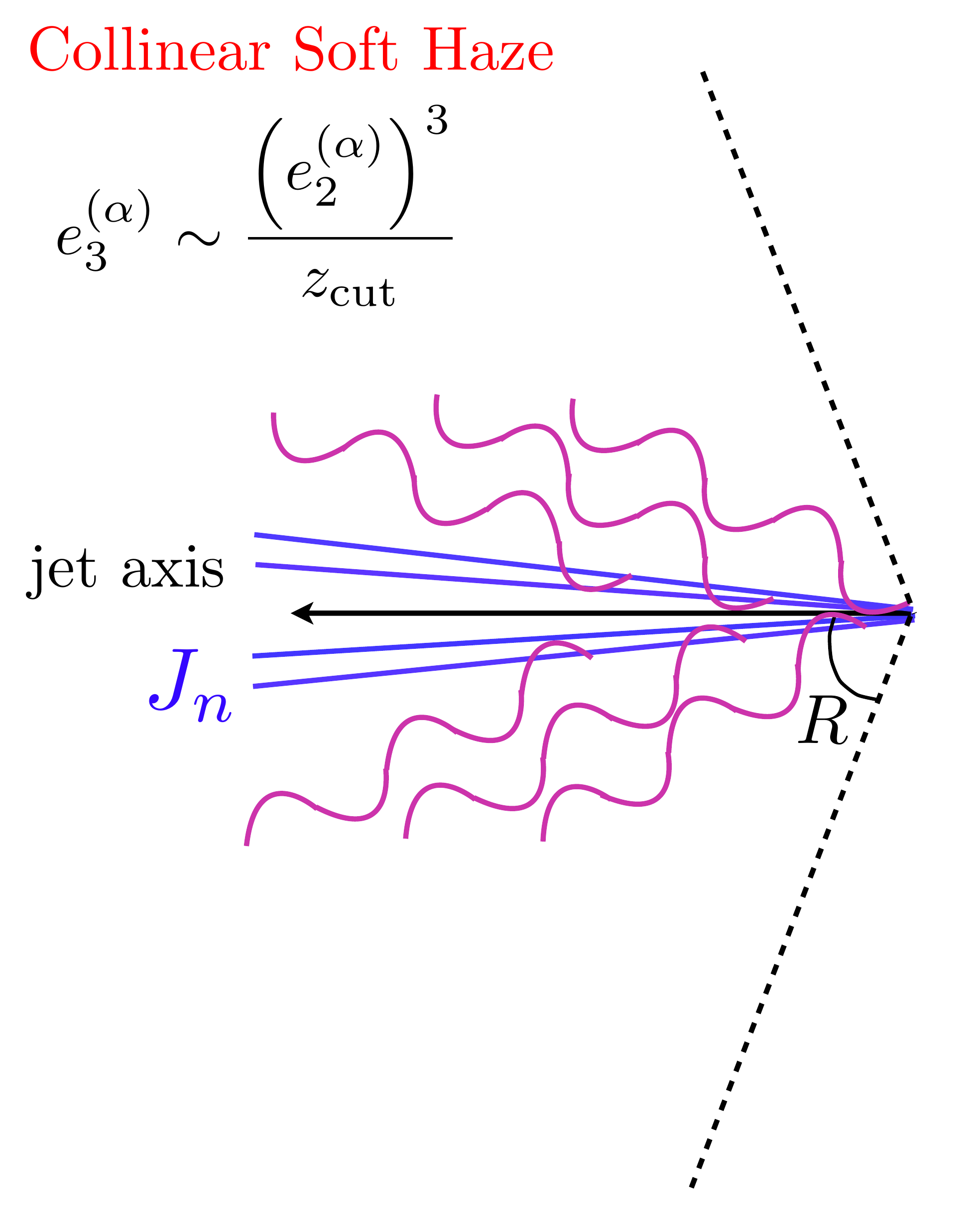}
\includegraphics[width=5cm]{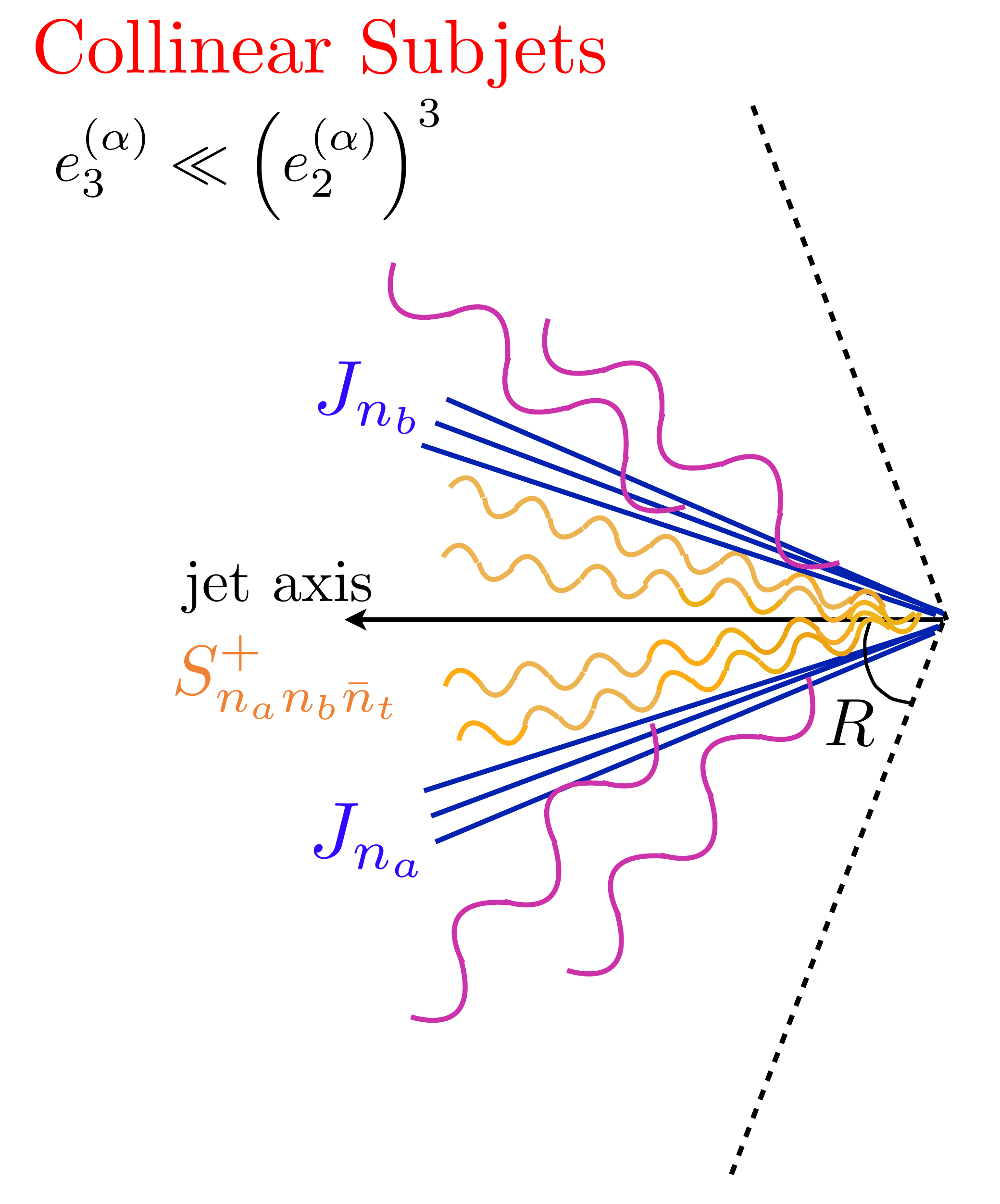}
\includegraphics[width = 5cm]{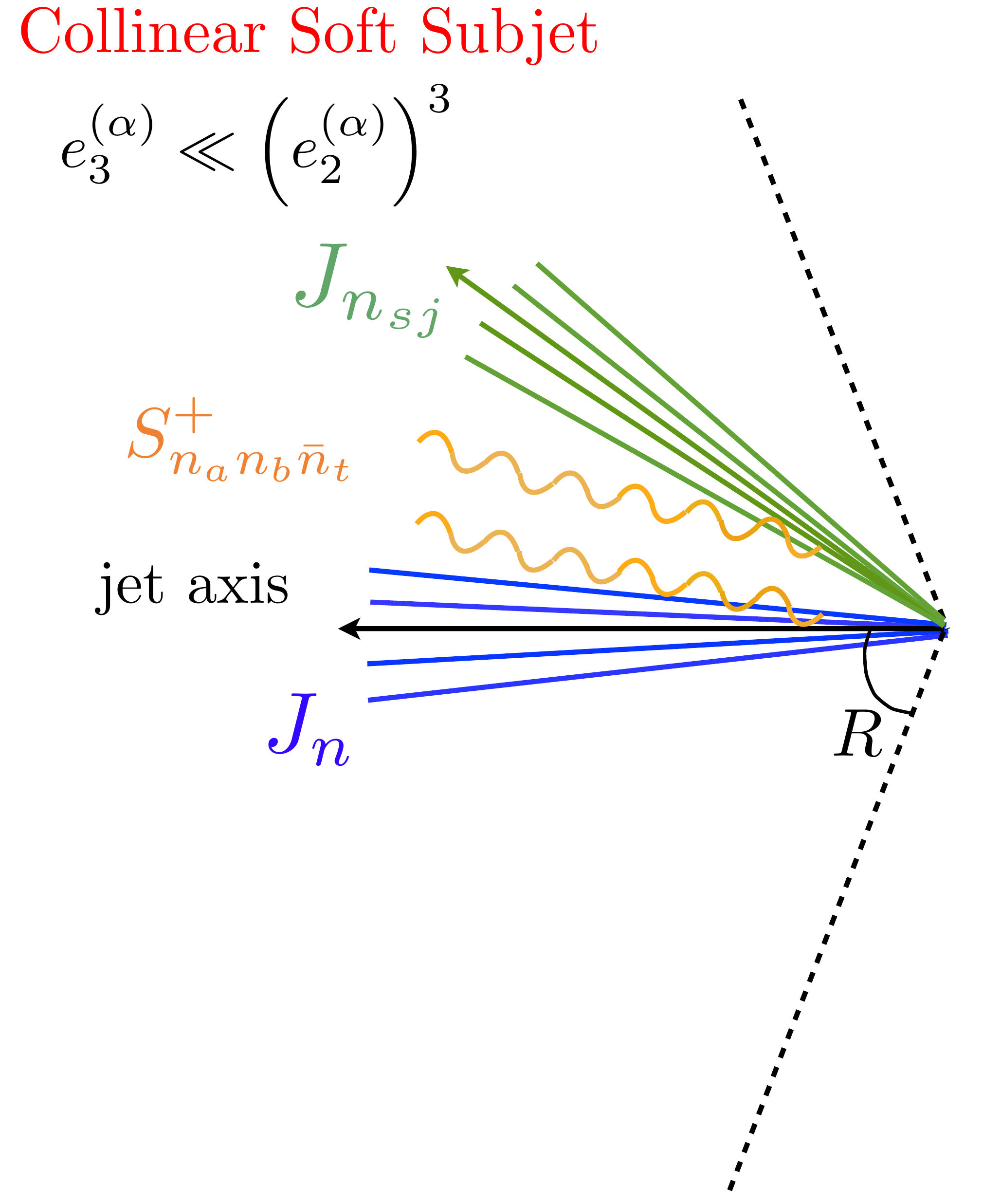}
\end{center}
\caption{Regions of interest for studying the two-prong substructure of a jet on which the soft drop grooming algorithm has been applied. (a) Collinear-soft haze region in which no subjets are resolved. (b) Collinear subjets with comparable energy and a small opening angle. (c) Collinear-soft subjet carrying a small fraction of the total energy, with $z_{cs} \sim z_{\text{cut}}$. 
}
\label{fig:subjets_summary}
\end{figure}

\begin{figure}
\begin{center}
\includegraphics[width=7cm]{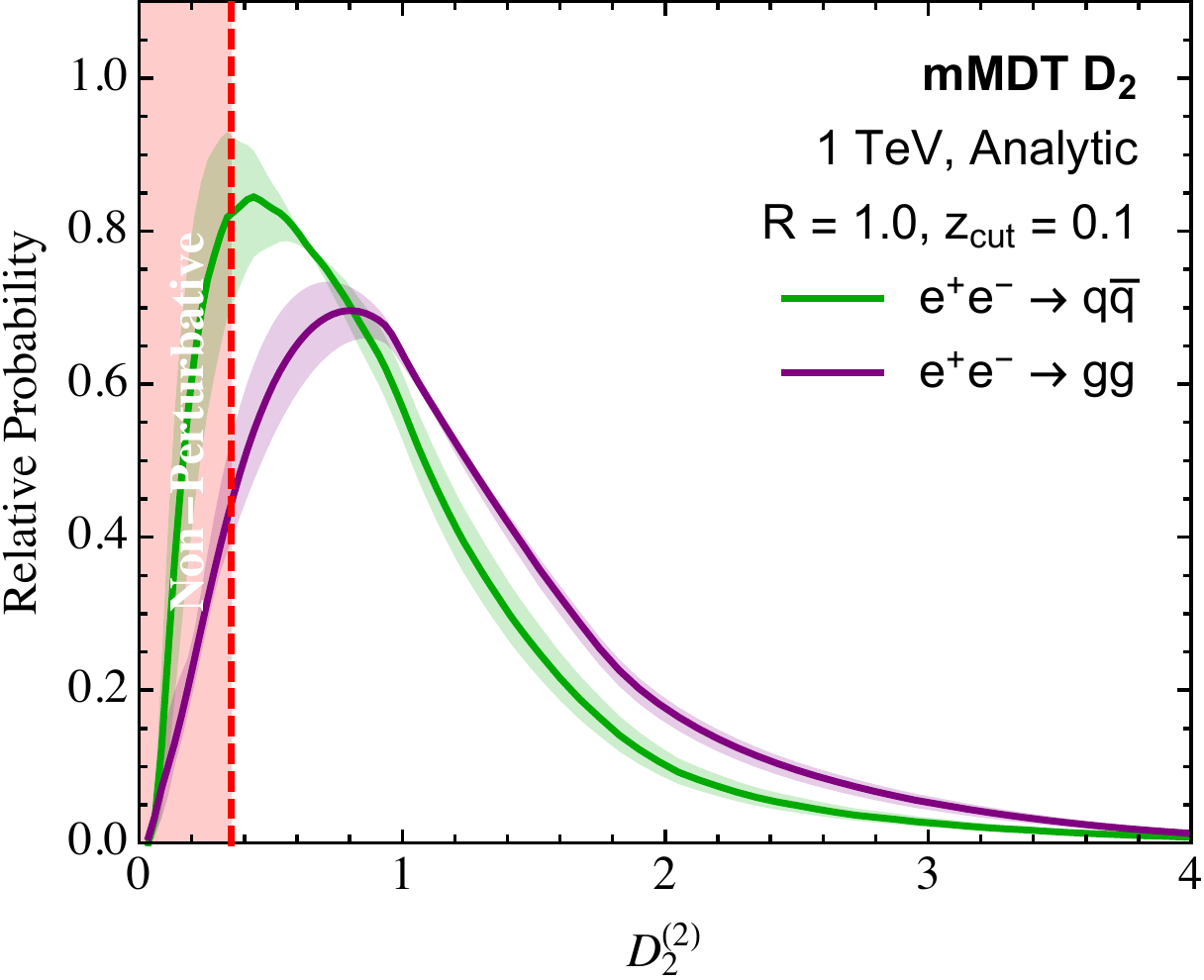}\qquad
\includegraphics[width=7cm]{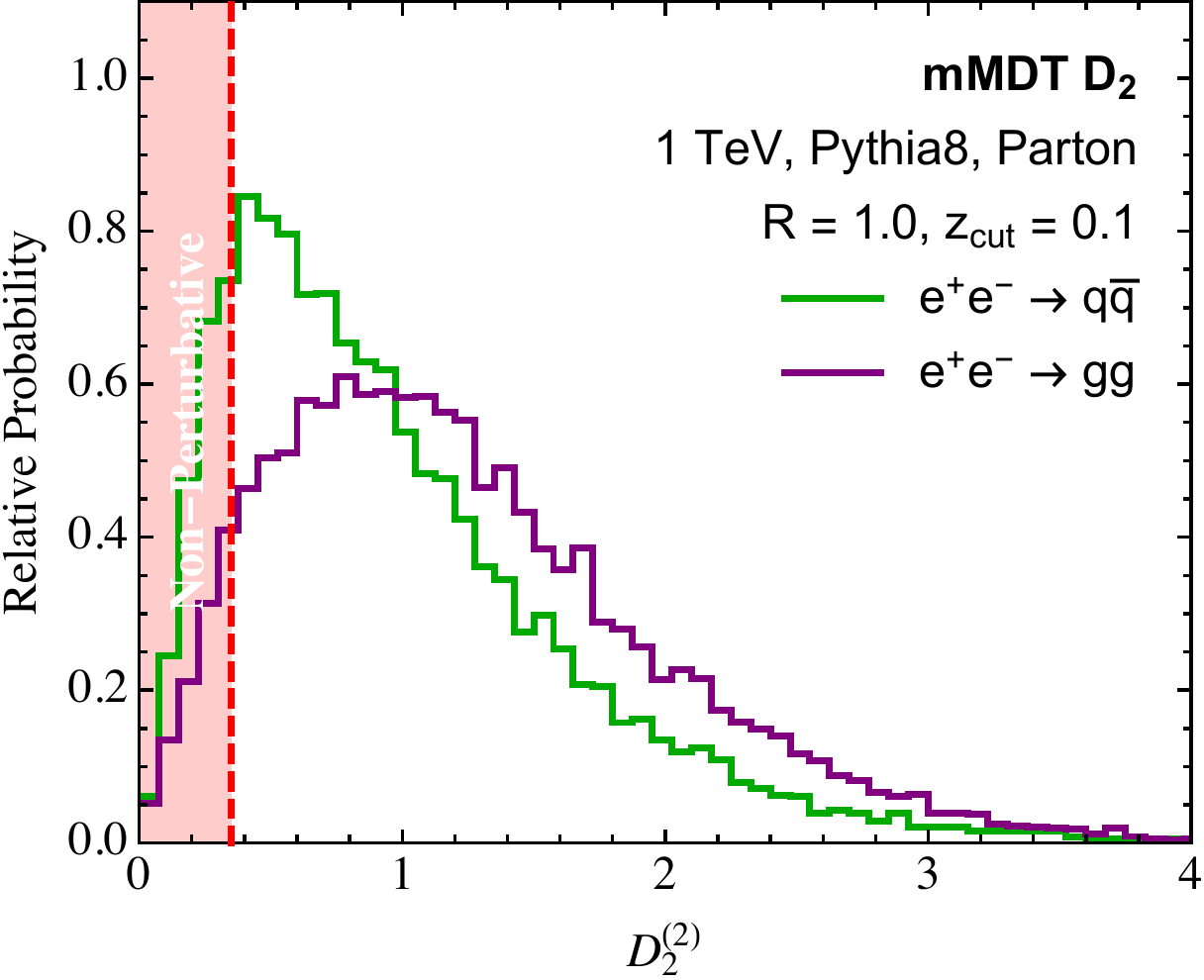}
\end{center}
\caption{\label{fig:D2_groomed}A comparison of the analytic $D_2$ distributions for both quark and gluons (left column) with parton shower Monte Carlo at parton level (right column), for $\zcut=0.1$. A mass cut of $m_J \in [80,100]$ GeV has been applied. Good agreement in the shape is observed. The analytic distribution were calculated at NLL accuracy, and matched to the leading order distribution. 
}
\label{fig:eecompplots}
\end{figure}

The utility of the energy-correlation shape variables\footnote{In prior literature, Refs. \cite{Larkoski:2013eya,Larkoski:2014gra,Larkoski:2015zka,Larkoski:2015kga,Larkoski:2017cqq}, they were often called energy correlation functions. But we will refrain from this here, as they are not actually a correlation function, though the moments of their distributions are related to correlation functions.} is their ability to cleanly separate out distinct phase-space regions. We have already noted that $e_{2}^{(2)}\ll 1$ selects for a narrow jet, but if we wish to understand the structure of a $1\rightarrow 2$ splitting process in QCD, we must dig deeper into the phase space. A prime example of how to do this is the ratio observable:
\begin{align}
    D_{2}^{(\beta)}&=\frac{e_{3}^{(\beta)}}{(e_{2}^{(\beta)})^3}\,.
\end{align}
While we will not reproduce the full analysis here, see Ref.\cite{Larkoski:2014gra}, when $D_{3}^{(\beta)}\ll 1$, we will have a jet with two-prongs. Otherwise, the jet will consist of predominately one central core of radiation. By imposing multiple constraints on $e_{2}^{\alpha},e_{2}^{\beta},e_{3}^{\gamma}$, for various exponents, we can further differentiate whether the jet has two collinear prongs, or a soft prong with a hard central prong. Radiation between the prongs is suppressed as long as we maintain the condition $D_{3}^{(\beta)}\ll 1$. These distinct configurations are illustrated in Fig.~\ref{fig:subjets_summary} in the case that we have also groomed the jet with the modified Mass Drop Tagger algorithm of Ref.~\cite{Dasgupta:2013ihk}, where wide-angle soft radiation at the edge of the jet, carrying an energy fraction less than $z_{\text{cut}}$ has been systematically removed. This enables writing down factorizations for each kinematic region, and the smooth interpolation between regions via subtractions of the overlaps. Such interpolations are more challenging using other substructure observables, for instance, N-subjettiness Refs.~\cite{Thaler:2010tr}. However, see Ref.~\cite{Dasgupta:2015lxh} for a wide ranging analysis of different multi-prong discriminants. In Fig.~\ref{fig:D2_groomed}, from \cite{Larkoski:2017cqq}, we give the distribution of $D_2^{(2)}$ for groomed jets.

Up till now, the main application of energy-correlation shape variables has been to tagging boosted objects coming out of a hadronic collision, relying on a first-principles analysis of the QCD process generating the decay patterns. As such, they continue to serve as a benchmark against which more sophisticated machine learning algorithms are tested. A worry always is that the machine learning algorithm will always be biased by the selection of the training data, so that any claimed improvement in performance over first-principle driven tagging variables should be accompanied by a first principles motivated analysis of the tagging procedure Ref.~\cite{Larkoski:2019nwj}.

\section{EEC and TEEC as Probes of TMD Physics in Deep Inelastic Scattering}

With the Electron-Ion Collider (EIC)  on the horizon and the very high resummed accuracy that can be achieved in the evaluation of  EEC and TEEC, we now turn to deep inelastic scattering (DIS) on nucleons and nuclei.    

\subsection{TEEC at the Electron-Ion Collider}

The TEEC observable can be generalized to DIS   by considering the transverse-energy and transverse-energy correlation between the lepton and hadrons in the final state~\cite{Li:2020bub}
\begin{align}~\label{eq:teec_dis}
    \text{TEEC} =&  \sum_{a} \int d\sigma_{lp\to l+a+X} \frac{ E_{T,l}  E_{T,a}}{E_{T,l} \sum_{i} E_{T,i}}  \delta(\cos\phi_{la}-\cos\phi)\, ,  
%    \nonumber \\ 
%    =& \sum_{a} \int d\sigma_{lp\to l+a+X} \frac{  E_{T,a}}{\sum_{i} E_{T,i}}  
%\delta(\cos\phi_{la}-\cos\phi) \, ,
\end{align}
where the sum runs over all the hadrons in the final state and $\phi_{la}$ is the  azimuthal angle between the final-state lepton $l$ and hadron $a$.

Taking DIS as an example, the underlying partonic Born process is 
$ e(k_1) + q(k_2) \to e(k_3) + q(k_4) $ and the first order non-trivial contribution to TEEC begins from one order higher. Similarly to TEEC in hadronic collisions, the cross section in the back-to-back limit is factorized into the convolution of a hard function, beam function, soft function, and jet function.
Specifically, up to leading power in SCET in terms of the variable $\tau = [1+\cos(\phi)]/2$ the cross section can be written as  
\begin{align}\label{eq:sing}
    \frac{d\sigma^{(0)}}{d\tau} =& \sum_{f}  \int\frac{d\xi  dQ^2 }{\xi Q^2}  Q_{f}^2 \sigma_0 \frac{p_T}{\sqrt{\tau}}\int \frac{db}{2\pi} e^{-2ib\sqrt{\tau} p_T}  B_{f/N}(b,E_2, \xi, \mu, \nu ) H(Q, \mu )
    \nonumber \\ & \times 
    S\left(b,\frac{n_2\cdot n_4}{2},\mu,\nu\right)J_{f}(b,E_4, \mu, \nu) \,,
\end{align}
where $\sigma_0= \frac{2 \pi \alpha^2}{Q^2}[1+(1-y)^2] $,  $b$ is the conjugate variable to $k_y$,  $Q^2$ is the invariant mass of the virtual photon, and $y=Q^2/(\xi s)$.  Four-vectors  $n_2$ and $n_4$ represent the momentum directions of the momenta $k_2$ and $k_4$, respectively. $E_2$ and $E_4$ are the energies of $k_2$ and $k_4$. $\nu$ is rapidity scale associated with the rapidity regulator for which we adopt the exponential regulator  introduced in Ref.~\cite{Li:2016axz}.

\begin{figure}[h]
    \centering
    \includegraphics[width=0.49 \textwidth]{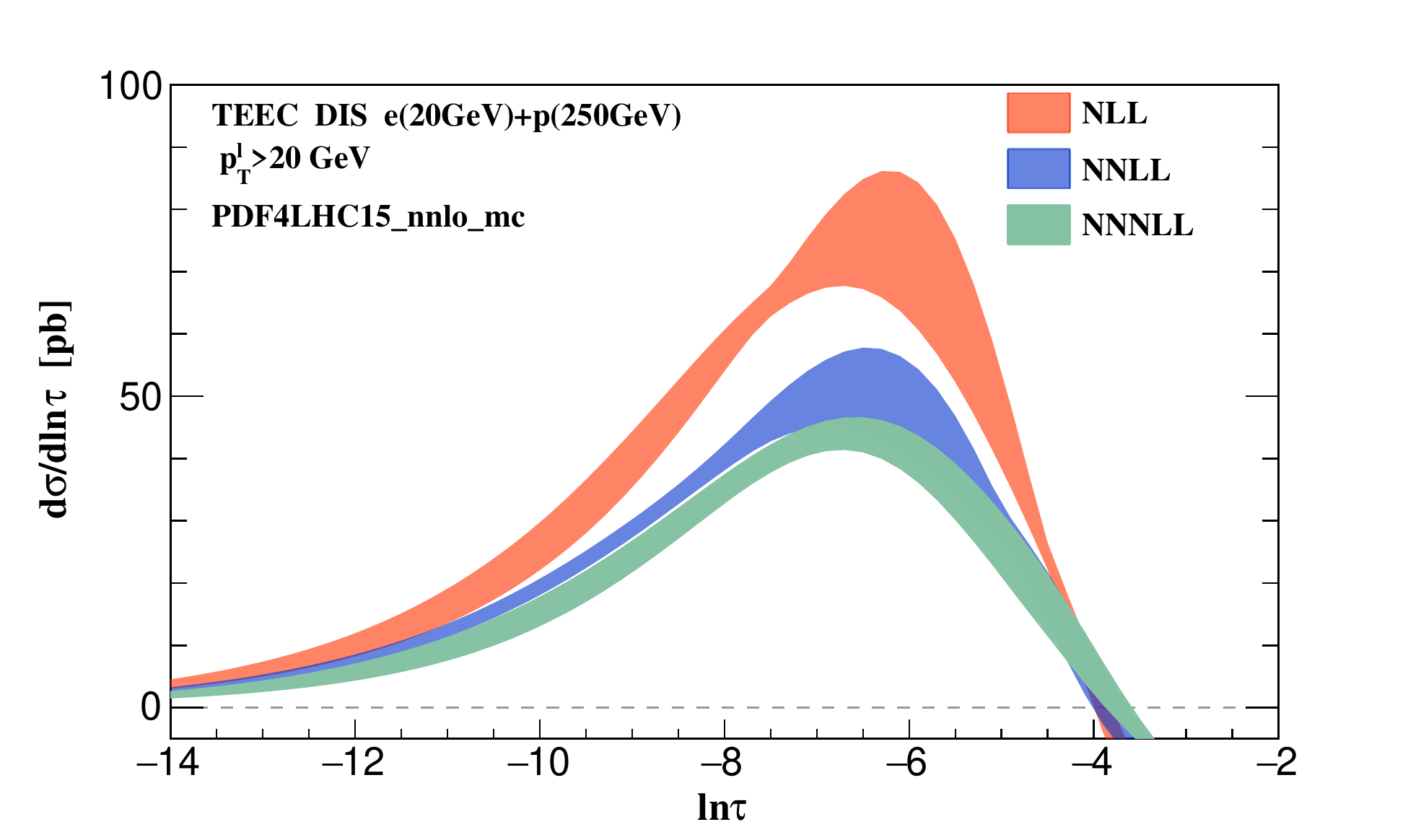}
    \includegraphics[width=0.49 \textwidth]{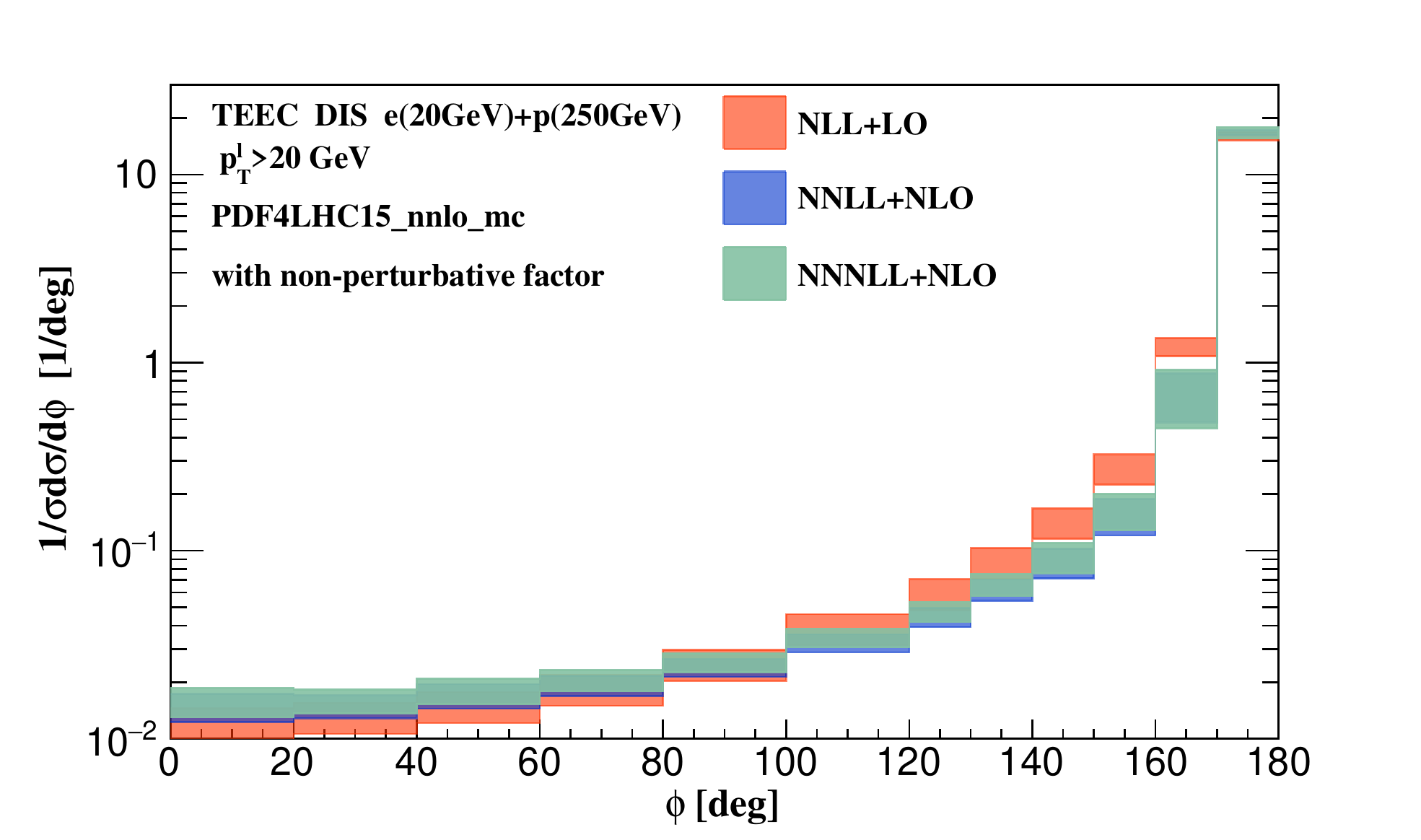}
    \caption{Left: resummed distributions in the back-to-back limit up to N$^3$LL accuracy. Note that results are not normalized by $\sigma$ in the $\tau$ interval shown.
 Right: TEEC $\phi$ distribution matched with a nonperturbative model. The orange, blue and green bands are the final predictions with scale uncertainties up to N$^3$LL+NLO. }
    \label{fig:TEEC}
\end{figure}

The connection to transverse momentum distribution physics is clear.   The TMD beam functions have been calculated up to three loops for quark beam functions \cite{Luo:2019szz,Ebert:2020yqt} and unpolarized gluon beam \cite{Ebert:2020yqt}. Two loops results are available for the linearly polarized beam function~\cite{Luo:2019bmw}. 
The jet function $J_f$ is defined as the second Mellin  moment of the matching coefficients of the TMD fragmentation function. The soft function $S$ is the same as the TMD soft function. In addition to the close connection to TMD physics, TEEC in DIS has the advantage that it can be computed to high accuracy. The left panel of \fig{TEEC} presents the resummed predictions at NLL, NNLL,  and N$^3$LL accuracy in the back-to-back limit with scale uncertainties~\cite{Li:2020bub}. Ref.~\cite{Li:2020bub} finds good perturbative convergence % \clee{[see my comment in the caption]}. 
There is about 30\% suppression in the peak region from NLL to NNLL, while it is about 5-6\% from NNLL to N$^3$LL. The reason is that these are absolute cross sections rather than ones normalized over a finite $\tau$ interval. The NLL uncertainty might also be underestimated. In general  the non-perturbative  (NP) corrections can be important in the infrared region and can be studied with the help of TEEC in DIS. The results  for the normalized TEEC $\phi$ distributions are shown in the right panel of Fig.~\ref{fig:TEEC}, where the non-perturbative Sudakov factor is also implemented~\cite{Li:2020bub}. The matching region is  chosen to be $160^{o}<\phi<175^{o}$ and for $\phi<160^{o}$ the distributions are generated by fixed-order calculations. The fixed-order predictions are calculated with $\mu_r=\mu_f=\kappa Q$ with $\kappa=(0.5,1,2)$. In the back-to-back limit, the predictions are significantly improved.

\subsection{EEC in the Breit Frame in DIS}

\begin{figure}[h]
\centering
\includegraphics[width=0.9\textwidth]{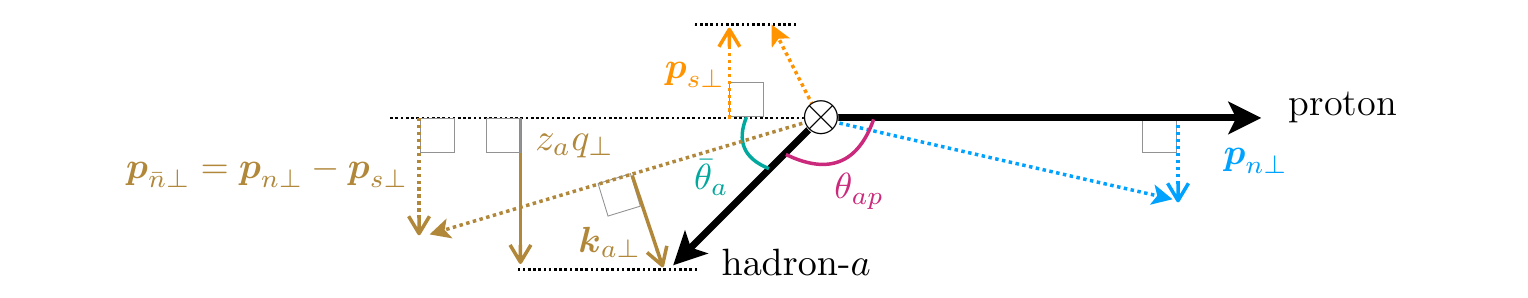}
 \caption{Illustration of the measurement of the transverse momentum $\boldmath{q}_{\perp}$ of the hadron-$a$ w.r.t. the proton axis in the Breit frame.}
\label{fig:measurment} 
\end{figure}

Measurements of QCD observables in DIS are often done in the Breit frame. 
Recently, a new definition of EEC in the Breit frame, which is a natural frame for the study of TMD physics~\cite{Collins:2011zzd}, was presented~\cite{Li:2021txc}.
In this frame, the target hadron moves along $\hat{z}$ and the  virtual photon  moves in the opposite direction. The Born-level process is described by the lepton-parton scattering $e+q_i\to e+q_f$, 
where the outgoing quark $q_f$ back-scatters in the direction opposite to the proton.  Hadronization of the struck quark will form a collimated spray of radiation close to the $- \hat{z}$ direction. On the other hand, initial state radiation and beam remnants are moving in the opposite direction close to the proton's direction of motion.  It is this feature of the Breit frame, which leads to the clean separation of target and current fragmentation that we utilize to construct the novel EEC observable in DIS.
The kinematics,  together with the contributions from the collinear and soft momenta to the transverse momentum of the hadron $q_\perp$ is illustrated in Figure~\ref{fig:measurment}.    

We denote the new event shape variable EEC$_{\text{DIS}}$ to avoid confusion with the conventional observable. Our definition reads,
\begin{equation} \label{eq:definition-DIS}
    \text{EEC}_{\text{DIS}}= \sum_a \int\, \frac{d\sigma_{e p \to e+ a+X}}{\sigma}\, z_a\, \delta(\cos\theta_{ap} - \cos\theta)\;,
\end{equation}
where
\begin{equation}\label{eq:weigth-DIS}
    z_a \equiv \frac{P\cdot p_a}{P\cdot (\sum_i p_i)}\;,
\end{equation}
and $p_a^{\mu}$ and $P^{\mu}$ are the momenta of the hadron $a$ and the incoming proton respectively. The angle $\theta_{ap}$ is the polar angle of hadron $a$, which is measured with respect to the incoming proton. Note that the asymmetric weight function, $z_a$, is Lorentz invariant and is suppressed for soft radiation and radiation close to the beam direction. Furthermore, this definition of EEC in the Breit frame naturally separates the contribution to the $\cos\theta$ spectrum from: i) wide angle soft radiation, ii) initial state radiation and beam remnants, and iii)  radiation from the hadronization of the struck quark. This unique feature makes the new observable in the back-to-back limit ($\theta \to \pi$) insensitive to experimental cuts on the particle pseudorapidity (in the Laboratory frame) due to detector acceptance limitations in the backward and forward regions, making the comparison of theory and experiment in this region even more accurate. This definition of EEC is spherically invariant, however, definitions that are fully Lorentz invariant and can be measured directly in any frame are also possible. 

\begin{figure}[htb]
    \centering
    \includegraphics[width=0.4\textwidth]{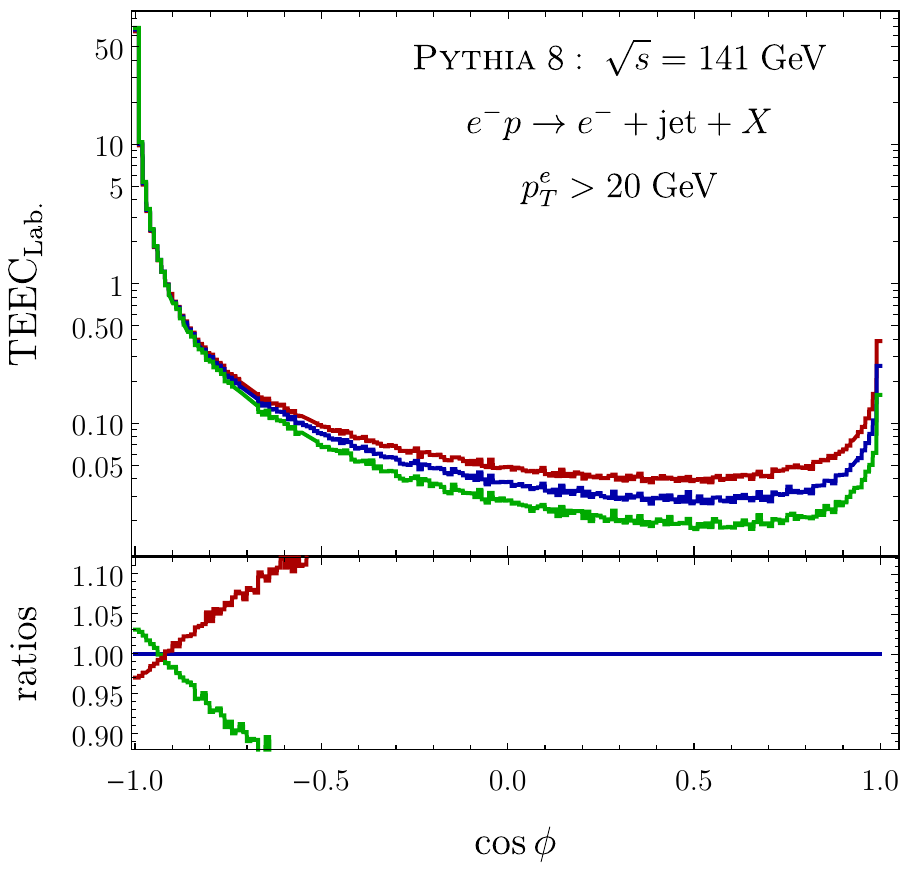}\;\;\hspace{0.4cm}
    \includegraphics[width=0.4\textwidth]{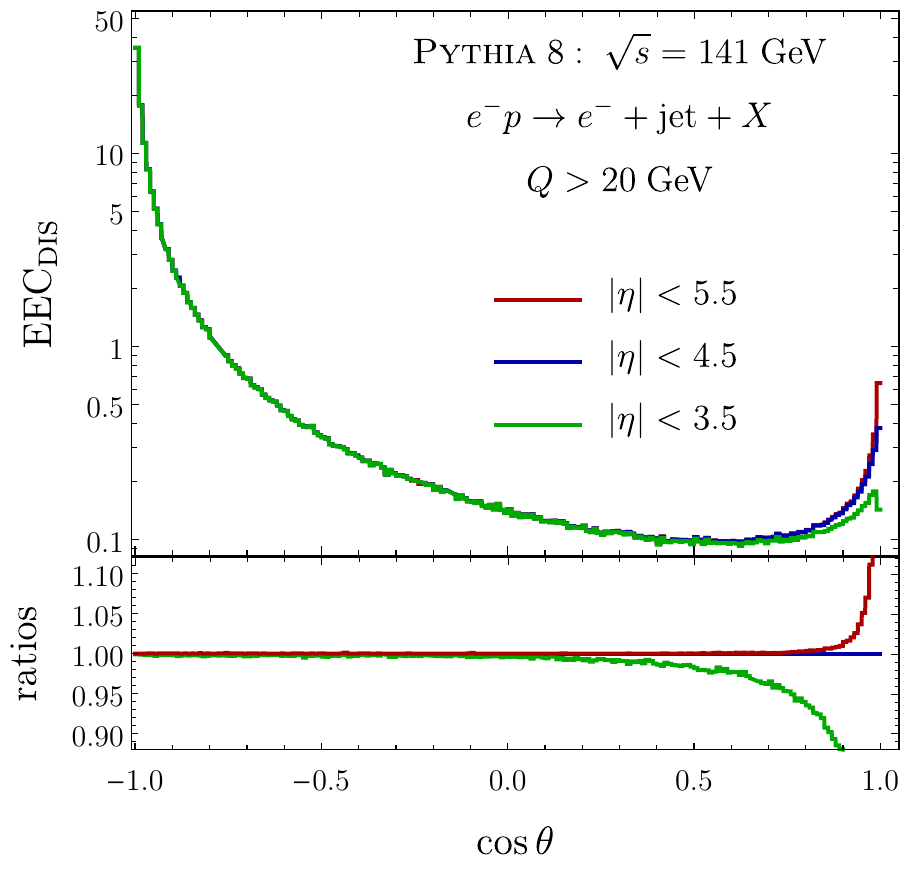}
    \vspace{-0.5cm}
    \caption{TEEC (left) and EEC (right) distributions from PYTHIA 8 with different rapidity cuts in the lab frame.  The ratio relative to the  
    $|\eta| < 4.5$ case is also shown.}
    \label{fig:pythia}
\end{figure}

To illustrate the reduced sensitivity of the new observable to kinematics, we present the TEEC$_{\rm Lab}$~\cite{Li:2020bub} and EEC$_{\rm DIS}$ distributions predicted by \textsc{Pythia} 8~\cite{Sjostrand:2007gs,Sjostrand:2014zea} in Fig.~\ref{fig:pythia}. The  red,  blue, and  green lines represent the results with pseudorapidity cuts $|\eta|<5.5$, $|\eta|<4.5$, and $|\eta|<3.5$ in the lab frame, respectively, which imitates detector limitations in the backward and forward regions. In order to compare the results with different pseudorapidity cuts,  all the distributions in Fig.~\ref{fig:pythia} are normalized by the event number with $|\eta|<5.5$.   Because TEEC measures the correlation between hadrons and the final state lepton in the lab frame, pseudorapidity cuts have an impact on the full $\cos\phi$ range, as shown in left panel of Fig.~\ref{fig:pythia}. EEC is defined as the correlation between the final state hadrons and incoming proton in the Breit frame, and the pseudorapidity cuts only remove particles in the forward region where the weighted cross section is small. In the backward region the distribution is independent on the pseudorapidity cuts.  Similar perturbative accuracy to the TEEC case can be achieved.  

\section{EEC in Dense Matter}

Jet and hadron correlations and jet substructure receive large in-medium corrections in reactions with nuclei. So far, these studies have been limited to heavy ion reactions~\cite{STAR:2002svs,ATLAS:2010isq,CMS:2017eqd,ATLAS:2012tjt,CMS:2013lhm,CMS:2020plq,Neufeld:2012df,Dai:2012am,Kang:2017xnc,Chien:2015hda,Li:2019dre}. Thus, EEC can also be used to shed light on the interaction between partons and cold QCD medium in electron-ion ($eA$) collisions. Ongoing theoretical and experimental design efforts aim to elucidate the physics opportunities with hadron and jet modification at the EIC and, very importantly, to ensure that  the detectors at this future facility have the capabilities to perform the necessary measurements, see e.g. ref.~\cite{Li:2020sru}. 

As recent calculations rely on the SCET framework, a natural choice to address  (T)EEC in $eA$ collisions  is the extension of the effective theory approach to include the interactions between partons and the background QCD medium mediated by Glauber gluons. Soft collinear effective theory with Glauber gluon interactions has provided a way to evaluate the contribution of in-medium parton  showers~\cite{Ovanesyan:2011xy,Kang:2016ofv} to a variety of observables in reactions with nuclei.  The most recent examples include the modification of jet cross sections and  jet substructure ranging from the jet splitting functions  to the jet 
charge~\cite{Li:2018xuv,Li:2017wwc,Li:2019dre}. To obtain predictions for the (T)EEC event shape observable in DIS on nuclei will require a computation of the contributions 
from parton branching  in strongly-interacting matter to the terms  in the master factorization formula.  This deserves careful consideration and can be one of our future research goals.

\section{Conclusions}
Due to its simplicity, the EEC is a natural candidate for precision studies of the dynamics of the strong interactions. 
In perturbative QCD the EEC is the event shape where we have the most analytic control, due to the analytic results at NLO % \cite{Dixon:2018qgp,Luo:2019nig} 
and the resummation at N$^3$LL$^\prime$ in the back-to-back limit.
% t \cite{Ebert:2020sfi}. 
Given its sensitivity to strong radiation, the EEC is also a natural candidate for the extraction of the strong coupling constant. Significant progress in this direction can be made by including the most recent perturbative results, although a much better control and understanding of non perturbative effects has to be obtained before achieving a level of accuracy comparable to the lattice results.

Moreover, the EEC/TEEC event shape observables can be studied in $e^+e^-$, $ep$ and $pp$ collisions, which provides a way to test the universality of QCD factorization in different colliding systems. These observables can also be used to study TMD physics, which is one of the most important goals of the EIC. 

Hadron colliders present a  more complex environment than $e^+e^-$ or $ep$ colliders, nevertheless, the simplicity of TEEC in $pp$ reactions shows that high accuracy can still be achieved. On one hand, this open the avenue for the rich LHC data to be combined with precision resummed QCD prediction to obtain precision measurement of standard model parameter, such as strong coupling constant, and various TMD functions. On the other hand, the simplicity in theoretical structure of EEC/TEEC make higher order calculation feasible.% by modern Feynman integral techniques. 
In turn these higher calculation can shed light on the structure of perturbation theory. Both of these lead to valuable addition to our understanding of QCD.

Finally, the EEC/TEEC observables can be generalized to $eA$,  $pA$, and $AA$ collisions. They can shed new light on the many-body QCD dynamics in reactions with nuclei, specifically  multi-parton  interactions and the formation of parton showers in matter. In these environments, precise extraction of transport properties of various forms of nuclear matter will greatly benefit from the high perturbative accuracy achieved in the baseline $ep$ and $pp$ reactions.

\acknowledgments
DN  is supported by the U.S. Department of Energy under Contract No. 89233218CNA000001 and by the LDRD program at LANL.
 GV is supported by the United States Department of Energy, Contract DE-AC02-76SF00515. 
 IV  is supported by the U.S. Department of Energy under Contract No. 89233218CNA000001 and by the LDRD program at LANL.
HXZ is supported by the National Natural Science Foundation of China under contract No. 11975200.

\bibliography{main}
\bibliographystyle{JHEP}
\end{document}